\documentclass[11pt,a4paper,nofootinbib,superscriptaddress]{revtex4-1}

\pdfoutput=1
\usepackage[utf8]{inputenc}
\usepackage{graphicx,bm,bbm}
\usepackage{amssymb,graphicx,epstopdf}
\usepackage{slashed,subfigure}
\usepackage{caption}
\usepackage{xcolor}
\usepackage{amsmath,mathtools}
\usepackage{epstopdf,dcolumn}
\usepackage{soul}
\usepackage{mathtools}
\allowdisplaybreaks
\usepackage[normalem]{ulem}
\usepackage{cancel}
\usepackage{xcolor}
\usepackage[colorlinks=true,linktocpage=true,linkcolor=blue,citecolor=blue]{hyperref}
\usepackage{xcolor}
\usepackage{braket}
\allowdisplaybreaks
\usepackage{enumitem}
\usepackage{mathrsfs}

\def\be {\begin{equation}}
\def\ee {\end{equation}}
\def\bea {\begin{eqnarray}}
\def\eea {\end{eqnarray}}
\def\bc {\begin{center}}
	\def\ec {\end{center}}
\def\nn {\nonumber}
\def\eps {\epsilon}
\def\gm {\gamma}

\def\mn {\mu\nu}
\def\al {\alpha}
\def\om{\omega}
\def\({\left(}
\def\){\right)}
\def\[{\left[}
\def\]{\right]}

\newcommand \Tr{\operatorname{\text{Tr}}}

\DeclareGraphicsExtensions{.jpg,.pdf,.eps}

\begin{document}

\title{Anisotropic tomography of heavy quark dissociation by using general propagator structure at finite magnetic field}
	
\author{Ritesh Ghosh}
\email{ritesh.ghosh@saha.ac.in}
\affiliation{
	Theory Division, Saha Institute of Nuclear Physics, HBNI, \\
	1/AF, Bidhannagar, Kolkata 700064, India}
\affiliation{
	Homi Bhabha National Institute, Anushaktinagar, \\
	Mumbai, Maharashtra 400094, India}
\author{Aritra Bandyopadhyay}
\email{aritrabanerjee.444@gmail.com}
\affiliation{Guangdong Provincial Key Laboratory of Nuclear Science, Institute of Quantum Matter, South China Normal University, Guangzhou 510006, China}
 \affiliation{Institut für Theoretische Physik, Universität Heidelberg, Philosophenweg 16, 69120 Heidelberg, Germany}
\author{Indrani Nilima}
\email{nilima.ism@gmail.com}
\affiliation{Department of Physics, Banaras Hindu University, Varanasi 221005, India}
 \author{Sabyasachi Ghosh}
 \email{sabya@iitbhilai.ac.in}
\affiliation{Indian Institute of Technology Bhilai, GEC Campus, Sejbahar, Raipur - 492015, Chhattisgarh, India}

\begin{abstract}
In this work we have explored the imaginary part of the Heavy Quark (HQ) potential and subsequently the dissociation of heavy quarkonia at finite temperature and magnetic field. With respect to earlier investigations on this topic, present work contain three new ingredients. First one is considering all Landau level summation, for which present work can be applicable in entire magnetic field domain - from weak to strong. Second one is the general structure of the gauge boson propagator in a hot magnetized medium, which is used here in heavy quark potential problem first time. 
Third one is a rich anisotropic structure of the complex heavy quark potential, which explicitly depends on the longitudinal and transverse distance. By comparing with earlier references, we have attempted to display our new contributions by plotting heavy quark potential tomography and dissociation probability at finite temperature and magnetic field.  

	\end{abstract}
	
	\maketitle 
		\tableofcontents
	\newpage
	
\section{Introduction}
	A lot of information is continuously being provided by Relativistic heavy-ion collisions (HIC) in the context of deconfined state of quark matter. Currently ongoing experimental programs has the aim to study the properties of quark matter at high temperatures, where it behaves as an weakly interacting gas of quarks and gluons. Recent surges for the new and exciting aspects of the quantum chromo dynamics (QCD) phase diagram has led to the idea of having a magnetized medium. It has been realized that the non central HIC produces a strong magnetic field~\cite{Kharzeev:2007jp,Skokov:2009qp} in the direction perpendicular to the reaction plane. At the RHIC and LHC energies, the strength of this magnetic field is estimated to be $B =m_{\pi}^2=10^{18}$ Gauss and $B =15m_{\pi}^2=1.5 \times10^{19}$ Gauss \cite{Kharzeev:2007jp,Skokov:2009qp} respectively, where $ m_{\pi} $ is the pion mass. Recent studies have also shown that such an extensive magnetic field might have existed in the early stages of Universe~\cite{Vachaspati:1991nm,Grasso:2000wj}.
	
	Several theoretical efforts~\cite{Kharzeev:2012ph,Miransky:2015ava} have been made to study the modification of the strongly interacting matter in presence of an external magnetic field. These studies include direct numerical investigations using lattice QCD~\cite{DElia:2010abb,DElia:2011koc,Bali:2012zg}
	and effective theoretical investigations using different methods
	including, e.g., anti-de Sitter/conformal field theory correspondence studies~\cite{Erdmenger:2011bw, Rougemont:2015oea,Gursoy:2016ofp}, perturbative studies~\cite{Alexandre:2000jc,Weldon:1982aq,Braaten:1989mz,Frenkel:1989br,Braaten:1991gm,Karmakar:2019tdp,Ghosh:2021knc,Bandyopadhyay:2017cle, Ghosh:2019kmf} and effective QCD model studies~\cite{Boomsma:2009yk,Gatto:2010qs,Agasian:2008tb,Andersen:2011ip,Fayazbakhsh:2012vr,Skokov:2011ib,Gorbar:2011jd,Gatto:2010pt,Mizher:2010zb,Andersen:2012dz,Preis:2012fh,Fraga:2012fs,Fraga:2012rr,Fraga:2008qn,Ferrari:2012yw,Ghosh:2022xtv,Fukushima:2012xw,dePaoli:2012cz,Kojo:2012js,Kojo:2013uua}. Subsequently those studies have anticipated many fascinating novel features of the magnetized medium. Although, to what extent those features  will be detected in the HIC experiments, is still not certain.
	
	In the present work we have focused on one important signature of quark matter and its behavior in presence of an arbitrary external magnetic field, i.e. heavy quarkonia. Because of their large mass and resistant behaviour towards thermal medium heavy quarkonia is considered as one of the most dynamic probes to study the characteristics of the quark matter. There are mainly two lines of theoretical approaches to determine quarkonium spectral functions, viz. the potential models~\cite{Matsui:1986dk,Karsch:1987pv,Mocsy:2008eg,Shuryak:2004tx,Wong:2004zr,Cabrera:2006wh}  and the lattice QCD studies~\cite{Satz:2006kba,Thakur:2012eb}. In lattice QCD simulation approach, one studies the spectral functions derived from Euclidean meson correlation~\cite{Alberico:2007rg}. Because of the decreasing temporal range at large temperature, construction of spectral functions is problematic and hence the results suffer from discretization effects and statistical errors and, thus, are still inconclusive. This is why potential models have been used widely to study the heavy quarkonia at finite temperatures. Pioneering studies to explore the heavy quarkonia and its dissociation at finite temperature using potential models has been done by Satz et. al.~\cite{Karsch:1987pv,Matsui:1986dk}. In Ref.~\cite{Matsui:1986dk}, they predicted a suppression of the bound state of $c {\bar c}$ pair, which is being caused by the shortening of the screening length for color interactions in the quark matter. 
	
	A very short lifetime ($\sim$ few fm/c) of the quark matter in HIC experiments further emphasizes the need to explore the effects of magnetic field on the properties of heavy quarkonia. There are also several studies in the literature which have explored the effect of magnetic field on the evaluation of quarkonia~\cite{Machado:2013rta,Guo:2015nsa,Marasinghe:2011bt,Yang:2011cz}. Modification of the heavy quark potential is one of the most important aspect of the theoretical up gradation required to study the properties of heavy quarkonia in a magnetized medium, which is studied recently by Refs.~\cite{Hasan:2017fmf,Singh:2017nfa}.
    The effect of a constant uniform magnetic field on the static quarkonium potential at zero and finite 
	temperature and on the screening masses have been studied in Refs.~\cite{Bonati:2016kxj} and ~\cite{Bonati:2017uvz}. For heavy quark diffusion phenomenology at finite magnetic field, see Refs.~\cite{Fukushima:2015wck,Das:2016cwd}.
	
	In this paper we will investigate the properties of heavy quarkonia at finite magnetic field using the most general structure of the gluon propagator in a hot magnetized medium. Several recent studies have presented various general structures of the fermion and gauge boson self energies vis-a-vis propagators at finite temperature and in presence of an external magnetic field~\cite{Shabad:2010hx,Hattori:2012je,Bordag:2008wp,Chao:2014wla,Mueller:2014tea,Das:2017vfh,Ayala:2018ina,Karmakar:2018aig,Ayala:2020wzl,Ayala:2021lor} using different independent tensor structures. For the present work we have chosen the effective gluon propagator in a hot and magnetized medium from Ref.~\cite{Karmakar:2018aig}. The medium modified heavy quark potential is the sum of both Coulombic and string terms~\cite{ Eichten:1974af} and it is directly dependent on the temporal component of the gluon propagator through the inverse of dielectric permittivity. For obtaining the imaginary parts of medium modified heavy quark potential we will extract the imaginary part of the resummed gluon propagator in terms of real and imaginary parts of gluon self energy form factors. Moreover, the form factors can be divided into fermionic and gluonic contributions and the magnetic field dependent contribution arises only from the fermionic contribution. Subsequently, this will give the imaginary part of the dielectric permittivity, which in turn will give the imaginary parts of the in-medium heavy quark potential~\cite{Agotiya:2008ie,Thakur:2012eb,Thakur:2013nia,Kakade:2015xua, Agotiya:2016bqr}.
	
	When we notice recent Refs.~\cite{Hasan:2017fmf,Singh:2017nfa} for the research topic on heavy quark potential at finite temperature and magnetic field, then we can find limitation of their application zone of magnetic field.  Ref.~\cite{Singh:2017nfa} can be applicable in the strong field limit, where lowest Landau level (LLL) approximation take dominant contribution, whereas Ref.~\cite{Hasan:2020iwa} is done in the weak field limit. In this regards, present work can be applicable to entire magnetic field domain from weak to strong as we are considering all Landau level summations.
	Unlike to Refs.~\cite{Singh:2017nfa,Hasan:2020iwa}, where they had neglected the Debye mass ($m_D$) independent terms in the calculation of the form factors, those terms will be automatically incorporated in our calculation. This contribution has been taken care by using general structure of gluon propagator at finite temperature and magnetic field. Apart from these two ingredients - (1) considering all Landau level summation and (2) considering general structure of gluon propagator at finite temperature and magnetic field, we also have shown the anisotropic structure of heavy quark potential, which can be naturally expected at finite magnetic field but ignored in earlier Refs.~\cite{Singh:2017nfa,Hasan:2020iwa}.  
	
	
	The paper is organized as follows. In section~\ref{sec1}, we will discuss the formalism used to execute this work. In subsection ~\ref{sec2} we discuss about the formalism of Heavy quark potential in presence of an external magnetic field. Subsections ~\ref{sec3} and ~\ref{sec4} deal with the formalism of imaginary part of the potential and evaluation of the real and imaginary parts of form factor $b(P)$ respectively. Moreover in subsection ~\ref{sec7},
	we will discuss about the final anisotropic expression of potential and decay width expression. Section ~\ref{sec9} refers to the results and discussions after which we conclude in section ~\ref{sec10}. Some zoom in calculations are provided in Appendix.

\section{Formalism}
\label{sec1}	
\subsection{Heavy quark potential in presence of an external magnetic field}	
\label{sec2}
To understand the melting of the quarkonia near crossover temperature, one needs to incorporate the non-perturbative effect in the heavy-quark potential. The Cornell potential consisting of the Coulomb and string-like part can describe the vacuum behavior of quarkonium bound state very well. In-medium behavior of the potential is not well-known in literature. There are several proposals to parametrize the real and imaginary part of potential and we would use one of them, by the virtue of which we can write down the in medium heavy quark potential in real space as~\cite{Lafferty:2019jpr,Thakur:2020ifi}
    \bea
	V(r)&=&\int \frac{d^3p}{(2\pi)^{3/2}} \bigg(e^{i\bold p \cdot \bold r}-1\bigg) \frac{V_{\text{Cornell}}(p)}{\eps(p)},
	\label{HQ_Vr}
	\eea
	where $\eps(p)$ is the dielectric permittivity which contains the medium information and $V_{\text{Cornell}}$ is the Cornell potential in momentum space, which is given by
	\bea
	V_{\text{Cornell}}(p)=-\sqrt{2/\pi}\frac{\al}{p^2}-\frac{4\sigma}{\sqrt{2\pi}p^4},
	\eea
	with $\al=C_F \al_s$, $C_F=(N_c^2-1)/2N_c$ and $\sigma$ is the string tension.
	
Inverse of dielectric permittivity $\eps(p)$ is related with the temporal component of the effective gluon propagator $D^{\mu\nu}$ by the definition~\cite{Singh:2017nfa}
\bea
\eps^{-1}(p)=\lim_{p_0\rightarrow0} p^2 D^{00}(P).
\eea
In presence of an external magnetic field one needs to consider appropriate modifications in the gluon propagator. The general structure of a gauge boson propagator in a hot magnetized medium is given in appendix \ref{appA} and from eq~\eqref{eff_prop} one can easily extract the temporal component as
\bea
D^{00}(P)&=& \frac{P^2-d}{(P^2-b)(P^2-d)-a^2}B^{00}(P),
\eea	
where $a(P),b(P)$ and $d(P)$ are the corresponding form factors whose explicit expressions are given in appendix \ref{appA}. Now in the vanishing limit of $p_0$, form factor $a(P)$ also vanishes~\cite{Alexandre:2000jc}. So we are not considering the form factor $a(P)$ in our case. Hence in our case the temporal component of the effective propagator can be further simplified as
\bea
D^{00}(P)&=& \frac{1}{(P^2-b)}B^{00}(P).
\label{eff_prop_simplified}
\eea

\subsection{Imaginary part of the potential}
Instead of going to real part of potential, we will focus directly on its imaginary part as it will be connected with
heavy quark dissociation probability, which is our matter of interest.
\label{sec3}
From Eq.~(\ref{HQ_Vr}), one can straightway extract the imaginary part of the in medium heavy quark potential as
\bea
\text{Im} V(r)&=&\int \frac{d^3p}{(2\pi)^{3/2}} \bigg(e^{i\bold p \cdot \bold r}-1\bigg) V_{\text{Cornell}}(p)\,\,\text{Im}\,\,\eps^{-1},\nn\\
&=&\int \frac{d^3p}{(2\pi)^{3/2}} \bigg(e^{i\bold p \cdot \bold r}-1\bigg) V_{\text{Cornell}}(p)\,\,p^2~\left(\lim_{p_0\rightarrow0}\text{Im}\, D^{00}(P)\right).
\label{imag_vr_initial}
\eea
In the spectral function representation one can evaluate the imaginary part of the effective gauge boson propagator ($D^{\mu\nu} = C_i \mathcal{P}_i^{\mu\nu}$, where $C_i$ are the form factors and $\mathcal{P}_i^{\mu\nu}$ are the projection operators) as~\cite{Weldon:1990iw}
\begin{align}
    \text{Im}\, D^{\mu\nu}(P\equiv \{p_0,{\bf p}\}) = - \pi \left(1+e^{-p_0/T}\right) \rho^{\mu\nu}(p_0,{\bf p}) ,
\end{align}
where $\rho^{\mu\nu}$ is the spectral function, represented as
\begin{align}
    \rho^{\mu\nu}(p_0,{\bf p}) = \frac{1}{\pi}~ \frac{e^{p_0/T}}{e^{p_0/T}-1}~\rho_i~ \mathcal{P}_i^{\mu\nu},
\end{align}
with $\rho_i$ being the imaginary parts of the respective form factors, i.e. $\rho_i = \text{Im}~C_i$. Using this approach, from equation~(\ref{eff_prop_simplified}) the imaginary part of the temporal component of the gluon propagator can be written in terms of the self energy form factor $b(P)$ as
\bea
\text{Im}\, D^{00}(p_0,\bold p)=-\pi(1+e^{-p_0/T})\times\frac{1}{\pi}\frac{e^{p_0/T}}{e^{p_0/T}-1} ~\frac{\text{Im}\, b}{(P^2-\text{Re}\,b)^2+(\text{Im}\, b)^2}~\frac{1}{\bar u^2},
\label{imag_D00_initial}
\eea 
where we have used $B^{00}(P) = \frac{1}{\bar{u}^2}$, with $\bar u=-\frac{\bold p^2}{p_0^2-\bold p^2}$. In the next subsection we will evaluate the real and imaginary part of the form factor $b(P)$.

\subsection{Evaluation of the real and imaginary parts of $b(P)$}	
\label{sec4}
The form factor $b(P)$ can be divided into quark and gluonic contributions as 
\bea
b(P) &=& b_q(P)+ b_g(P) =-\frac{p_0^2-p^2}{p^2}\bigg[\Pi^{00}_q(P) + \Pi^{00}_g(P)\bigg],
\label{b}
\eea
where $\Pi^{00}_{q/g}$ are quark/gluonic parts of the temporal component of the one loop gluon self energy in a hot magnetized medium. 

\subsubsection{Gluonic contribution}
\label{sec5}
Since gluons are not affected by magnetic field, the gluonic contribution of the one loop self energy is similar to the $B=0$ case~\cite{Singh:2017nfa}, i.e.
\bea
\Pi^{00}_g&=&m_{Dg}^2\bigg[1-\frac{p_0}{2p}\ln\bigg|\frac{p_0+p}{p_0-p}\bigg|+i\pi \frac{p_0}{2p}~\Theta(p^2-p_0^2)\bigg],
\eea
where $m_{Dg}^2=\frac{g^2 T^2 N_c}{3}$ and $\Theta$ is the step-function. This implies in the $p_0\rightarrow 0$ limit we can write down the real and imaginary part of the form factor $b_g(P)$ as
\bea
\lim_{p_0\rightarrow0}\text{Re}~b_g(P) &=& m_{Dg}^2, \label{real_bg}\\
\lim_{p_0\rightarrow0}\text{Im}~b_g(P) &=& \lim_{p_0\rightarrow0} m_{Dg}^2 ~\frac{\pi p_0}{2p}~\Theta(p^2). \label{imag_bg}
\eea

\subsubsection{Fermionic contribution}
\label{sec6}
\begin{center}
	\begin{figure}[tbh]
		\begin{center}
			\includegraphics[scale=0.40]{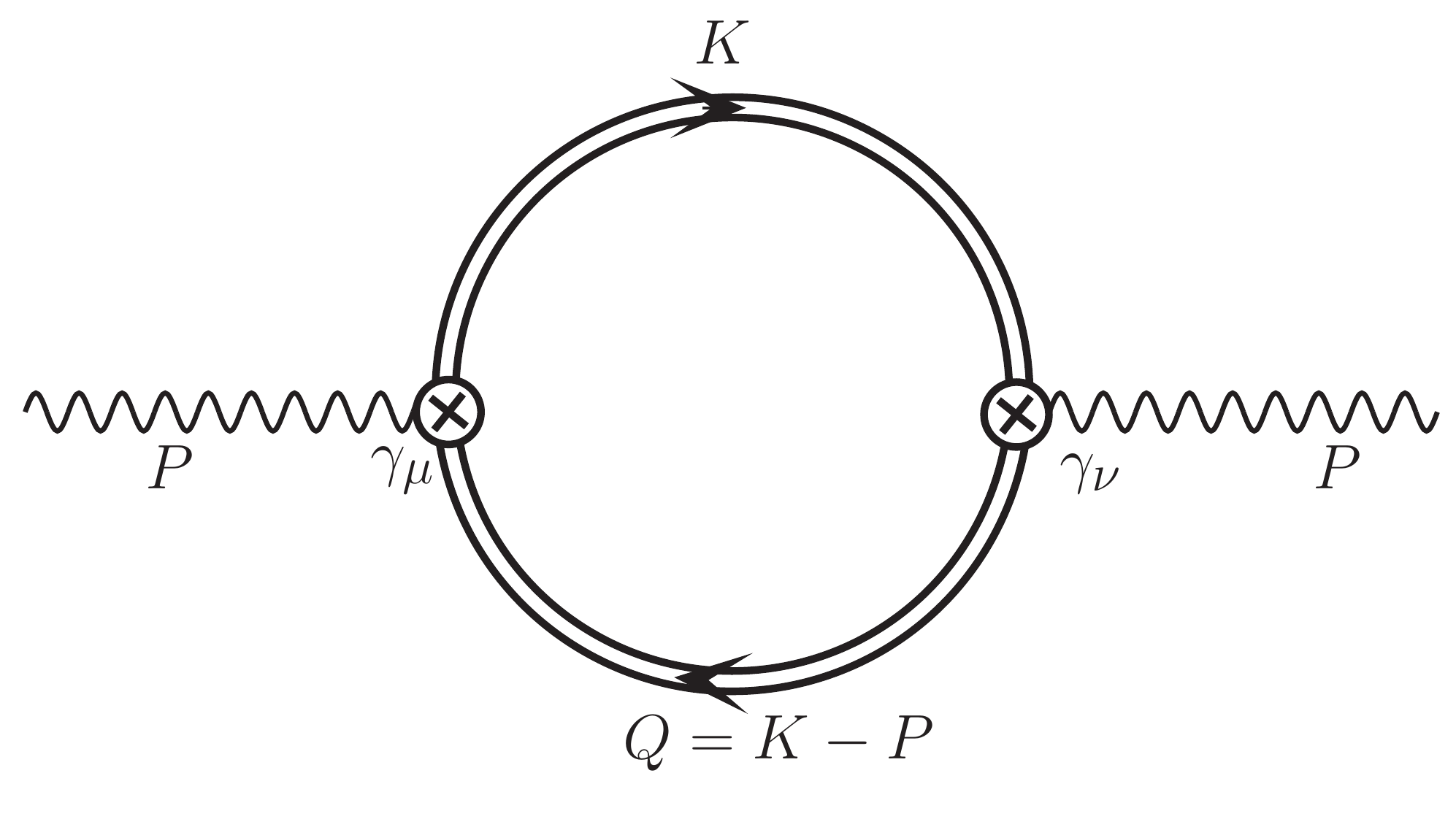}
			\caption{ One loop gluon self-energy.}
			\label{cvT_B0}
		\end{center}
	\end{figure}
\end{center}
To evaluate the fermionic contribution we are going to review the one loop gluon self-energy calculation from quark-antiquark loop in presence of arbitrary magnetic field. Qurak-antiquark are affected by magnetic field, so the propagator should be modified in presence of the magnetic field.
Translation invariant part of the fermion propagator $G(t,\bold{r})$ can be written in mixed coordinate-momentum space as~\cite{Miransky:2015ava,Wang:2020dsr} 
\bea
G(t,\bold{r})=\int \frac{d\om dk_z}{(2\pi)^2} e^{i k_z z-i\om t} G(\om,k_z,\bold{r}_\perp),
\eea
where 
\bea
G(\om,k_z,\bold{r}_\perp)=i \frac{e^{-\bold{r}^2_\perp/(4d_f^2)}}{2\pi d_f^2} \sum_{l=0}^{\infty}\frac{D_l(\om,k_z,\bold{r}_\perp)}{\om^2-k_z^2-m_f^2-2 l |eB|}.
\label{G_om}
\eea
In this case, the magnetic filed $\bold B$ is in the z-direction and the vector potential is given by Landau gauge i.e. $\bold A \equiv (-By,0,0)$. The numerator of eq.~\eqref{G_om} is given by
\bea
D_l(\om,k_z,\bold{r}_\perp)=(\om \gm^0-k_z \gm^3+m)\bigg[P_+\, L_l\bigg(\frac{\bold{r}^2_\perp}{2d_f^2}\bigg)+P_-\, L_{l-1}\bigg(\frac{\bold{r}^2_\perp}{2d_f^2}\bigg)\bigg]-\frac{i}{d_f^2}(\bold{r}_\perp \cdot \gm_\perp) L_{l-1}^1\bigg(\frac{\bold{r}^2_\perp}{2d_f^2}\bigg)~,
\nn\\
\eea
where $P_{\pm}=\frac{1}{2}[1\pm i\, \text{sign}(q_f B)\, \gm^1 \gm^2]$ are spin projectors and $d_f=\frac{1}{\sqrt{|q_f B|}}$. $L_n^\alpha(x)$ are the generalized Laguerre polynomials and $L^\alpha_{-1}(x)=0$ by definition.
So the gauge-boson self-energy can be written as
\bea
&&\Pi^{\mn}_q(i\om_m,\bold{p})\nn\\
&=&g^2 T\frac{1}{2}\sum_{n=-\infty}^{\infty}\int \frac{d k_z}{2\pi} d^2 \bold{r}_\perp e^{-i \bold{r}_\perp \cdot \bold{p}_\perp} \Tr\bigg[\gm^\mu G(i\om_n,k_z,\bold{r}_\perp)\gm^\nu G(i\om_n-i\om_m,k_z-p_z,-\bold{r}_\perp)\bigg].
\label{Pimunuq_1}
\eea
Here fermionic and bosonic Matsubara frequencies are $\om_n=(2n+1)\pi T$ and $\om_m=2m\pi T$ respectively. Eq.~(\ref{Pimunuq_1}) can be further simplified as
\bea
\Pi^{\mn}_q(i\om_m,\bold{p})=-g^2 T\frac{1}{2}\sum_{n=-\infty}^{\infty}\int \frac{d k_z}{2\pi} d^2 \bold{r}_\perp e^{-i \bold{r}_\perp \cdot \bold{p}_\perp} \frac{e^{- \bold{r}_\perp^2/(2c_f^2)}}{(2\pi d_f^2)^2}\sum_{l=0}^{\infty}\sum_{l'=0}^{\infty} \frac{1}{k_0^2-E_{l,k_z}} \frac{1}{q_0^2-E_{l',q_z}}S^{\mn},
\eea
where the fermionic energies are defined as $E_{l,k_z,f}=\sqrt{m_f^2+k_z^2+2 l|q_fB|}$. Here we define $S^{\mn}$ as the trace
\bea
S^{\mn}=\Tr[\gm^\mu D_l(i\om_n,k_z,\bold{r}_\perp)\gm^\nu D_l'(i\om_n-i\om_m,k_z,\bold{r}_\perp)].
\eea
Since we are only interested to find the temporal component of the one loop self energy, we can straightway put $\mu=\nu=0$ to get
\bea
\Pi^{00}_q(i\om_m,\bold p)&=& -g^2 T\frac{1}{2}\sum_{n=-\infty}^{\infty}\int \frac{d k_z}{2\pi} d^2 \bold{r}_\perp e^{-i \bold{r}_\perp \cdot \bold{p}_\perp} \frac{e^{- \bold{r}_\perp^2/(2d_f^2)}}{(2\pi d_f^2)^2}\sum_{l=0}^{\infty}\sum_{l'=0}^{\infty} \frac{1}{k_0^2-E_{l,k_z}} \frac{1}{q_0^2-E_{l',q_z}}\nn\\
&\times&\bigg\{2(L_l L_{l'}+L_{l-1}L_{l'-1})(k_0 q_0+k_3 q_3+m_f^2)+\frac{4  \bold{r}_\perp ^2}{d_f^4} L_{l-1}^1L_{l'-1}^1\bigg\}\nn\\
&=& -g^2 T \frac{1}{2}\sum_{n=-\infty}^{\infty}\int \frac{d k_z}{2\pi} \sum_{l=0}^{\infty}\sum_{l'=0}^{\infty} \frac{1}{k_0^2-E_{l,k_z}} \frac{1}{q_0^2-E_{l',q_z}} \frac{1}{4\pi^2 d_f^4}\nn\\
&\times&\bigg\{4\pi d_f^2(X_{l,l'}+X_{l-1,l'-1})(k_0 q_0+k_3 q_3+m_f^2)+8\pi  X_{l-1,l'-1}^1\bigg\}\nn\\
&=&-\frac{g^2}{4\pi^2}\frac{1}{2}\sum_{f=u,d} \frac{1}{d^4_f}\int \frac{dk_z}{2\pi}\sum_{l,l'=0}^{\infty}\,\,\sum_{s_1,s_2=\pm 1} \frac{n_F(E_{l,k_z,f})-n_F(s_1E_{l',q_z,f})}{4s_1E_{l,k_z,f}E_{l',q_z,f}}\nn\\
&\times&\frac{1}{is_2\, \om_m+E_{l,k_z,f}-s_1 E_{l',q_z,f}}(I_{1,f}+I_{2,f}),
\eea
where the functions $I_{1,f}$ and $I_{2,f}$ are defined as 
\bea
I_{1,f}&=&4\pi d_f^2 \bigg(s_1 E_{k_z,l}E_{q_z,l'}+k_z q_z+m_f^2\bigg)\bigg[X_{l,l'}+X_{l-1,l'-1}\bigg],\nn\\
I_{2,f}&=& 8\pi X^1_{l-1,l'-1}.
\eea
 The associated function $X_{m,n}$ and $X_{m,n}^1$ are defined in Appendix~\ref{app:B}. Finally we can write down the real part of the temporal component of the self energy in the limit of $p_0\rightarrow 0$ as
\bea
 \Pi^{00}_q(p_0,\bold p)\bigg|_{p_0\rightarrow 0} &=& -\frac{g^2}{8\pi^2}\sum_{f=u,d} \frac{1}{d^4_f}\int \frac{dk_z}{2\pi}\sum_{l,l'=0}^{\infty}\,\,\sum_{s_1,s_2=\pm 1}\nn\\ &&\left(\frac{n_F(E_{l,k_z,f})-n_F(s_1E_{l',q_z,f})}{4s_1E_{l,k_z,f}E_{l',q_z,f}}
\frac{I_{1,f}+I_{2,f}}{E_{l,k_z,f}-s_1 E_{l',q_z,f}}\right)\Bigg|_{p_0\rightarrow 0},
\label{real_bq}
\eea

Now to find out the imaginary part of the self-energy, we need to perform analytic continuation to the real value of gluon energy. By replacing $i\om_m\rightarrow p_0+i\eps$, the imaginary part of the temporal component of the gluon self-energy is given by,
\bea
\text{Im}\Pi^{00}_q(p_0,\bold p)&=& \frac{g^2}{4\pi}\frac{1}{2}\sum_{f=u,d} \frac{1}{d^4_f}\int \frac{dk_z}{2\pi}\sum_{l,l'=0}\sum_{s_1,s_2=\pm 1} \frac{n_F(E_{l,k_z,f})-n_F(s_1E_{l',q_z,f})}{4s_1s_2E_{l,k_z,f}E_{l',q_z,f}}\nn\\
&\times&\delta(s_2\, p^0+E_{l,k_z,f}-s_1 E_{l',q_z,f})(I_{1,f}+I_{2,f}).
\eea
In the limit of our interest, i.e. $p_0 \rightarrow 0$, only two delta function will contribute i.e. for $s_2=\pm 1$ when $s_1=1$. We can write the above equation as 
\bea
\text{Im}\Pi^{00}_q(p_0,\bold p)\bigg|_{p_0\rightarrow 0}&=&\frac{1}{2} \frac{g^2}{4\pi}\sum_{f=u,d} \frac{1}{d^4_f}\int \frac{dk_z}{2\pi}\sum_{l,l'=0}\sum_{s_2=\pm 1} \frac{1}{4s_2E_{l,k_z,f}E_{l',q_z,f}} \frac{\partial n_F(E_k)}{\partial E_k}s_2 p_0\nn\\
&\times&\delta(E_{l,k_z,f}- E_{l',q_z,f})(I_{1,f}+I_{2,f})\nn\\
&=&\frac{1}{2}\frac{2g^2}{4\pi}p_0\sum_{f=u,d} \frac{1}{d^4_f}\int \frac{dk_z}{2\pi}\sum_{l,l'=0} \frac{I_{1,f}+I_{2,f}}{4E_{l,k_z,f}E_{l',q_z,f}} \frac{\partial n_F(E_k)}{\partial E_k} 
\delta(E_{l,k_z,f}- E_{l',q_z,f}).
\eea
Now we use the following property of the Dirac delta function
\bea
\delta (f(x))=\sum_n \frac{\delta(x-x_n)}{|\frac{\partial f(x)}{\partial x}|_{x=x_n}},
\eea
with $x_n$ as the zeros of the function $f(x)$, to obtain the solutions for $k_z$ as
\bea
k_{z0}= \frac{2(l'-l)|q_fB |+p_z^2}{2p_z}.
\eea
Using the value of $k_{z0}$ subsequently we obtain the explicit expressions of fermionic energies as
\bea
E_{k_z,l}\bigg|_{k_z=k_{z0}}&=&\sqrt{m_f^2+2 l |q_f B|+\bigg(\frac{p_z^2+2(l'-l)|q_fB|}{2p_z}\bigg)^2},\\
E_{q_z,l'}\bigg|_{k_z=k_{z0}}&=&\sqrt{m_f^2+2 l' |q_f B|+\bigg(\frac{p_z^2-2(l'-l)|q_fB|}{2p_z}\bigg)^2}.
\eea
Hence, finally we can write 
\bea
\text{Im}\Pi^{00}_q(p_0,\bold p)\bigg|_{p_0\rightarrow 0} &=&-\beta\frac{2g^2}{4\pi}\frac{1}{2}p_0\sum_{f=u,d} \frac{1}{d^4_f} \frac{1}{2\pi}\sum_{l,l'=0} \frac{I_{1,f}+I_{2,f}}{4\,p_z \,E_{l,k_z,f}} n_F(E_k)(1-n_F(E_k))\bigg|_{kz=k_{z0}}
\label{imag_bq}
\eea

So, the real and imaginary parts of the form factor $b_q$ in the limit of $p_0\rightarrow 0$ are respectively given in Eq.~(\ref{real_bq}) and Eq.~(\ref{imag_bq}).

\subsection{Final expression of Imaginary part of potential and Decay width}
\label{sec7}
Using Eqs.~(\ref{real_bg}), (\ref{imag_bg}), (\ref{real_bq}) and (\ref{imag_bq}) in Eq.~(\ref{imag_D00_initial}) within the limit $p0 \rightarrow 0$, we find
\bea
\text{Im}\, D^{00}(\bold p)&=&-2\frac{1}{(p^2+\text{Re}\, b(p_0=0,\bold p))^2}\times\nn\\
&&\bigg(\frac{\pi T m_{Dg}^2}{2p}
-\frac{g^2}{4\pi}\sum_{f=u,d} \frac{1}{d^4_f} \frac{1}{2\pi}\sum_{l,l'=0} \frac{I_{1,f}+I_{2,f}}{4\,p_z \,E_{l,k_z,f}} n_F(E_k)(1-n_F(E_k))\bigg|_{kz=k_{z0}}\bigg),\label{imD00}
\eea
where 
\bea
\text{Re}\, b(p_0=0,\bold p)&=&\text{Re}\, b_q(p_0=0,\bold p)+\text{Re}\, b_g(p_0=0,\bold p),\nn\\
&=&\text{Re}\, \Pi^{00}_q(p_0=0,\bold p)+m_{Dg}^2.
\eea

As the expression for $\text{Im}~D^{00}$ is an explicit function of $p_z$ and $p_\perp$, so we need to accordingly break up the phase space due to anisotropy of the external magnetic field along the `$z$' direction. By doing that, Eq.~(\ref{imag_vr_initial}) will be transformed into
\begin{align}
\text{Im} V(r_\perp,z) &=-\int \frac{p_\perp dp_\perp dp_z d\phi_p}{(2\pi)^{3/2}} \bigg(e^{ip_\perp(x \cos \phi_p+y \sin \phi_p)+i zp_z}-1\bigg) \bigg(\sqrt{2/\pi}\frac{\al}{p^2}+\frac{4\sigma}{\sqrt{2\pi}p^4}\bigg)p^2\text{Im}\, D^{00}(p_z,p_\perp),\nn\\
&=-\int \frac{p_\perp dp_\perp  }{(2\pi)^{3/2}}\int_0^\infty dp_z\,4\pi \bigg(J_0(p_\perp r_\perp)\cos{ zp_z}-1\bigg) \bigg(\sqrt{2/\pi}\frac{\al}{p^2}+\frac{4\sigma}{\sqrt{2\pi}p^4}\bigg)p^2\text{Im}\, D^{00}(p_z,p_\perp).\label{imag_vr_final}
\end{align}
Eq.~(\ref{imag_vr_final}) is our final expression for the imaginary part of the heavy quark potential.

We will now use the imaginary part of the potential to calculate decay width $(\Gamma)$. So using the first-order time-independent perturbation theory, decay width$(\Gamma)$ can be estimated from the given equation~\cite{Thakur:2013nia,Singh:2017nfa}
\bea
\Gamma(T,B) = - \int d^3{\bf{r}}\, \left|\Psi({{r}})\right|^2{\rm{Im}}~V({\hat{r};T,B})~,
\label{Gamma}
\eea
Here $\psi(r)$ is the Coulombic wave function for the ground state is given by
\bea
\psi(r)=\frac{1}{\sqrt{\pi a_0^3}}e^{-r/a_0},
\eea
where $a_0=2/(m_Q \al)$.
Substituting the imaginary part of equation given in ~(\ref{imag_vr_final}) into (\ref{Gamma}) we estimate the decay width for given temperature and magnetic field. We would discuss the decay width of two quarkonia, $J/\psi$ (the ground state of charmonium, $c\bar c$ ) and $\Upsilon$ (bottomonium, $b\bar b$). 

\section{Results}
\label{sec9}

In this section we will discuss our results about the imaginary part of the HQ potential and the decay rate. For our present study we have chosen $N_c = 3$, $N_f = 2$ and the strong running coupling constant $g$ as 
\begin{eqnarray}
\label{gs}
g^2(T) = \frac{24 \pi^2}{\left(11N_c-2 N_{f}\right)\ln \left(\frac{2\pi T}{\Lambda_{\overline{\rm MS}}}\right)},
\end{eqnarray}
with $\Lambda_{\overline{\rm MS}} = 0.176$ GeV~\cite{Haque:2014rua}. 
We also want to mention here, that there are recent studies which explore the thermo-magnetic behavior of the strong coupling $g$~\cite{Ayala:2014uua,Ferrer:2014qka,Ayala:2018wux,Farias:2014eca}, which will be interesting to incorporate in future works.
We have taken the value of string tension as $\sigma=0.174\,\,\text{GeV}^2$~\cite{Burnier:2015nsa}.
Considering the anisotropy encountered in our studies, throughout the results section we will discuss two cases, i.e. with varying $z$ for a fixed $r_\perp$ and vice versa.

\begin{figure}[tbh]
\begin{center}
\includegraphics[scale=0.5]{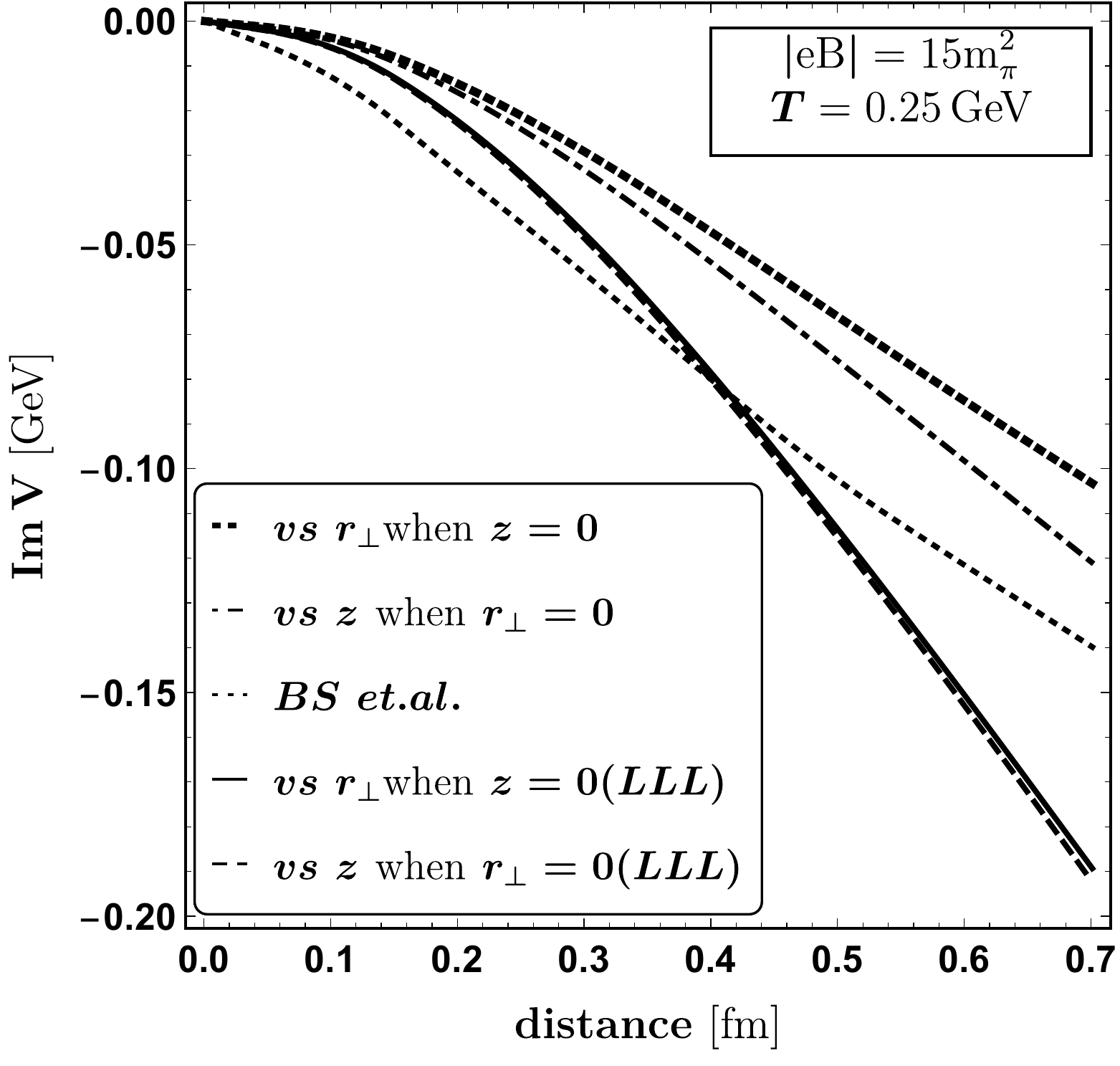}
\includegraphics[scale=0.51]{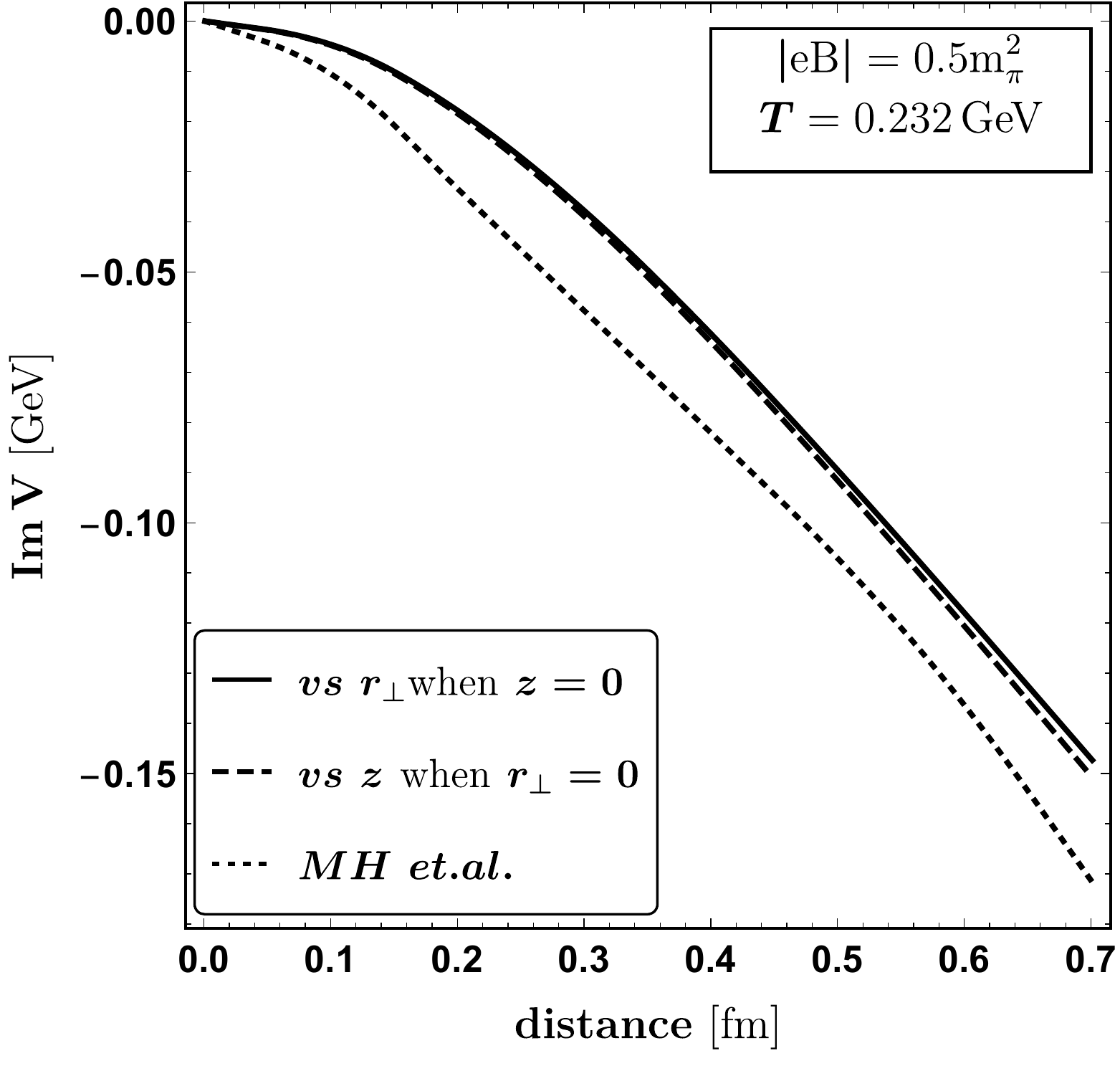}
\caption{Variation of ${\rm Im}~V$ with distance. We have shown two plots comparing with two recent results from Ref.~\cite{Singh:2017nfa} (left panel) and Ref.~\cite{Hasan:2020iwa} (right panel) which requires certain fixed values of magnetic field ($eB$) and temperature ($T$), as depicted in the plot. We have considered both the cases, i.e. with vanishing $r_\perp$ and with vanishing $z$.}
\label{imagV_comparison}
\end{center}
\end{figure}

As mentioned in the introduction, recently several studies have explored the heavy quark potential in a magnetized medium restricting themselves to limiting cases involving strong or weak magnetic field approximations. In this context, present work can be applicable in entire domain of magnetic field from weak to strong as we are considering all Landau level summation. Therefore, we have started our numerical presentation from Fig.~\ref{imagV_comparison}, where we have compared our result with two such recent results, strong field or LLL approximated result from Singh et.al.~\cite{Singh:2017nfa} and weak field or perturbatively expanded result from Hasan et.al.~\cite{Hasan:2020iwa}. Since our present calculation has captured the anisotropic outcomes of magnetic field, so potential become function of $r_\perp$ and $z$ but earlier Refs.~\cite{Singh:2017nfa,Hasan:2020iwa} provide isotropic potential in terms of $r$ only. Hence, the comparison will not be very straight forward.    
In the left panel, we have compared our anisotropic results for $r_\perp = 0$ and for $z=0$ with the LLL approximated results from Ref.~\cite{Singh:2017nfa} which shows noticeable difference between them. 
One of the source of this difference is that our results carry all Landau level summation but Ref.~\cite{Singh:2017nfa} is LLL approximation.
To find other sources of difference, we have generated our LLL approximated results which also differs from that of Ref~\cite{Singh:2017nfa}. Origin of this difference between the two LLL approximated results can be traced back to the structure of the coefficient function $b$ where we have made no approximations unlike Ref~\cite{Singh:2017nfa}, where they have neglected the Debye mass ($m_D$) independent terms. Also anisotropic and isotropic structures are another level of differences. Similar difference can again be observed in the right panel of fig.~\ref{imagV_comparison} where we have compared our general results for both $r_\perp = 0$ and $z=0$ with that of an weakly approximated one from Ref~\cite{Hasan:2020iwa}. Hence, the left and right panels of Fig.~\ref{imagV_comparison} indicate that our results in weak and strong fields both limits can not merged with earlier estimations~\cite{Singh:2017nfa,Hasan:2020iwa} because of general structure of magnetized gluon propagator and anisotropic structure of potential, considered in the present work.    


\begin{figure} 
\begin{center}
\includegraphics[scale=0.5]{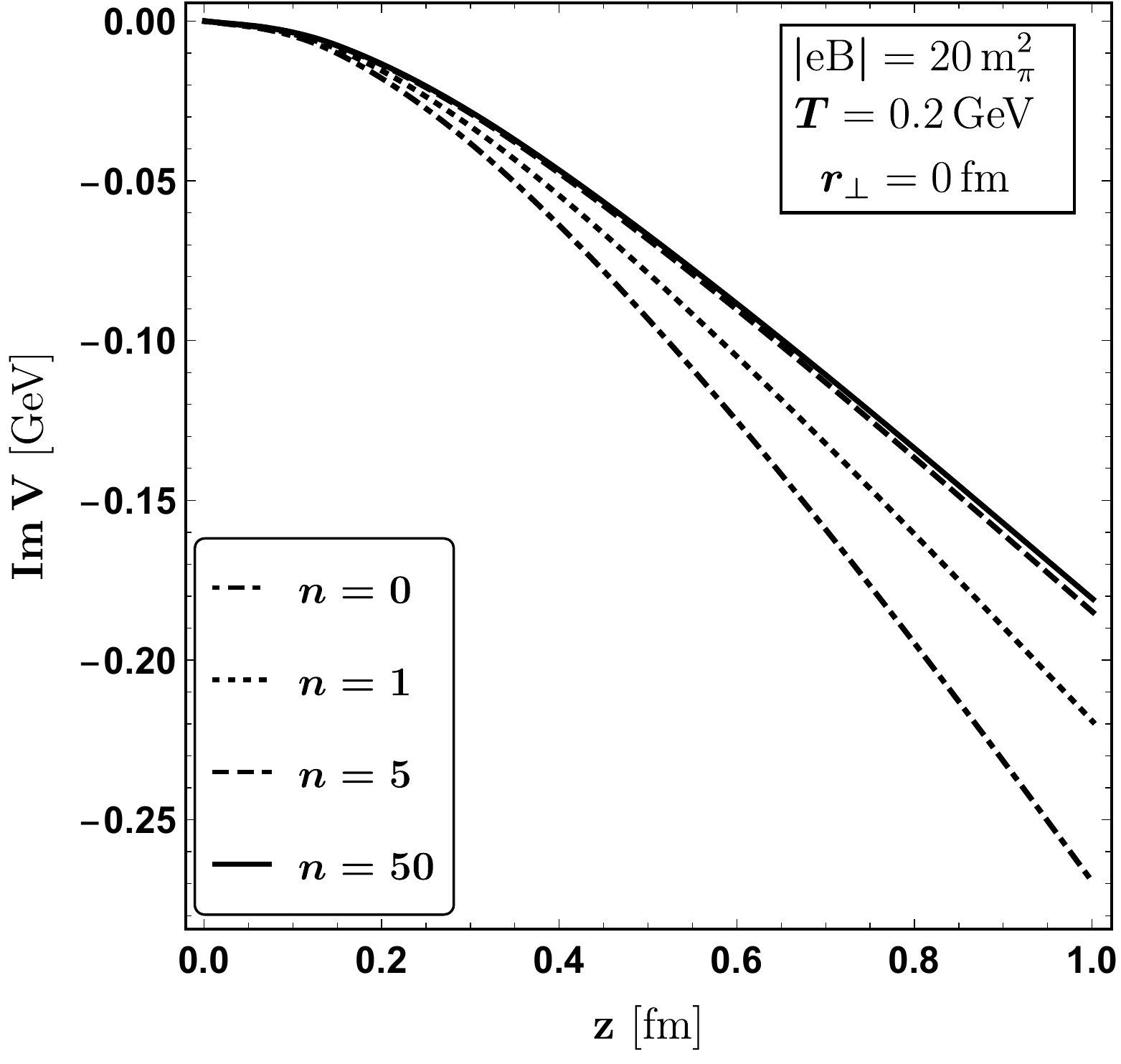}
\includegraphics[scale=0.5]{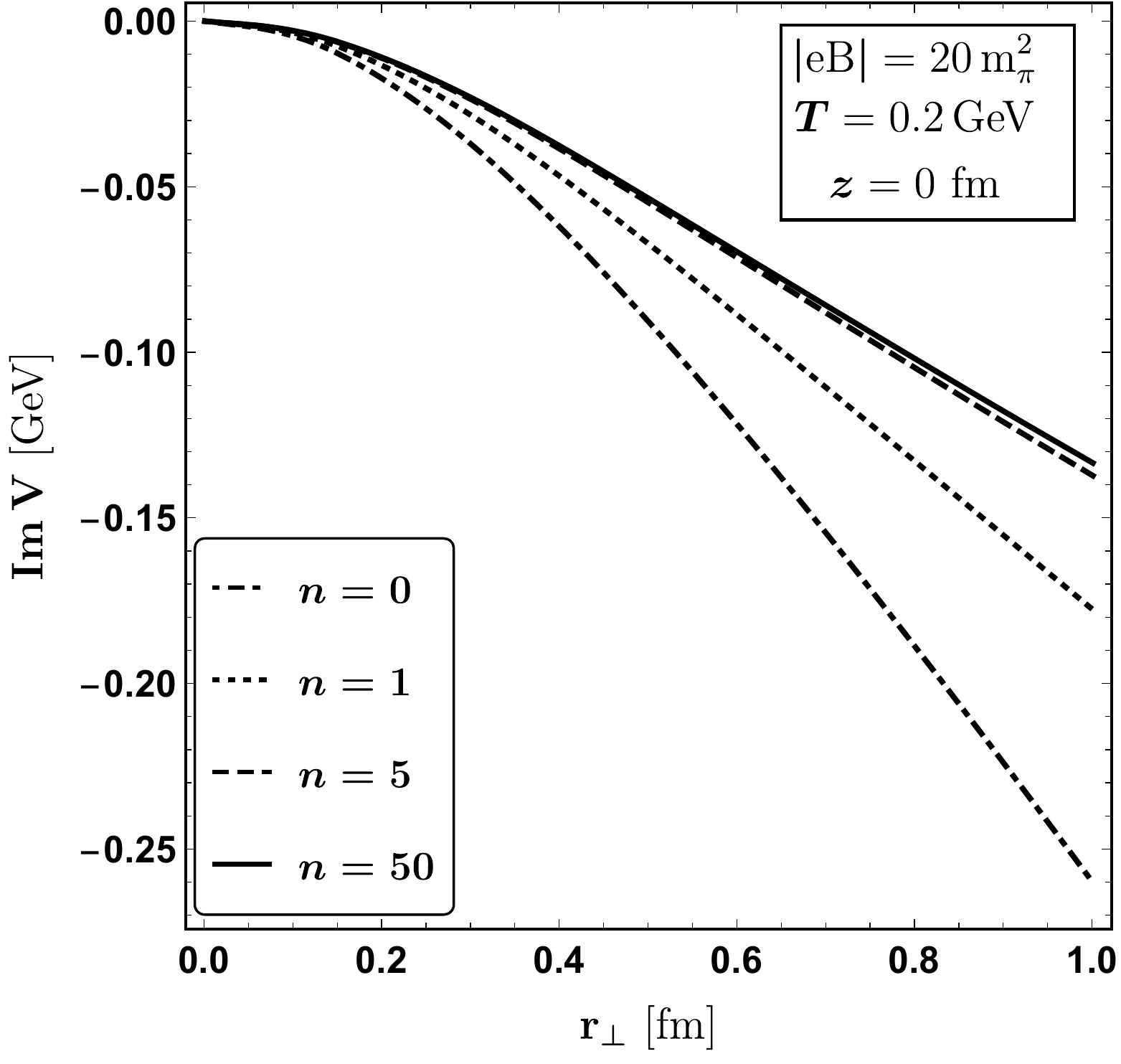}
\caption{Variation of ${\rm Im}~V$ with distance, i.e. with $z$ for vanishing $r_\perp$ (left panel) and with $r_\perp$ for vanishing $z$ (right panel), is shown considering different number of Landau levels where we show the difference between the LLL approximated result and the full result.}
\label{imagV_LLLvsfull}
\end{center}
\end{figure}

To further emphasise the deficiency of the LLL approximation, in fig.~\ref{imagV_LLLvsfull} we have plotted the variation of the imaginary part of the HQ potential with distance for various increasing values of the Landau levels and compared them with respect to the LLL approximated result. Again, we have shown two different cases in two panels of fig.~\ref{imagV_LLLvsfull}, left panel showing $r_\perp=0$ case and right panel showing $z=0$ case. For both the cases one can identify that the LLL approximated result is hugely overestimating the values for the imaginary part of the HQ potential, whereas with increasing values of the number of Landau levels $n$, the gap with the full result is getting diminished. We have considered $n=50$ as full results since we notice that the values are not changing beyond $n=10$.


\begin{figure}[tbh]
\begin{center}
\includegraphics[scale=0.5]{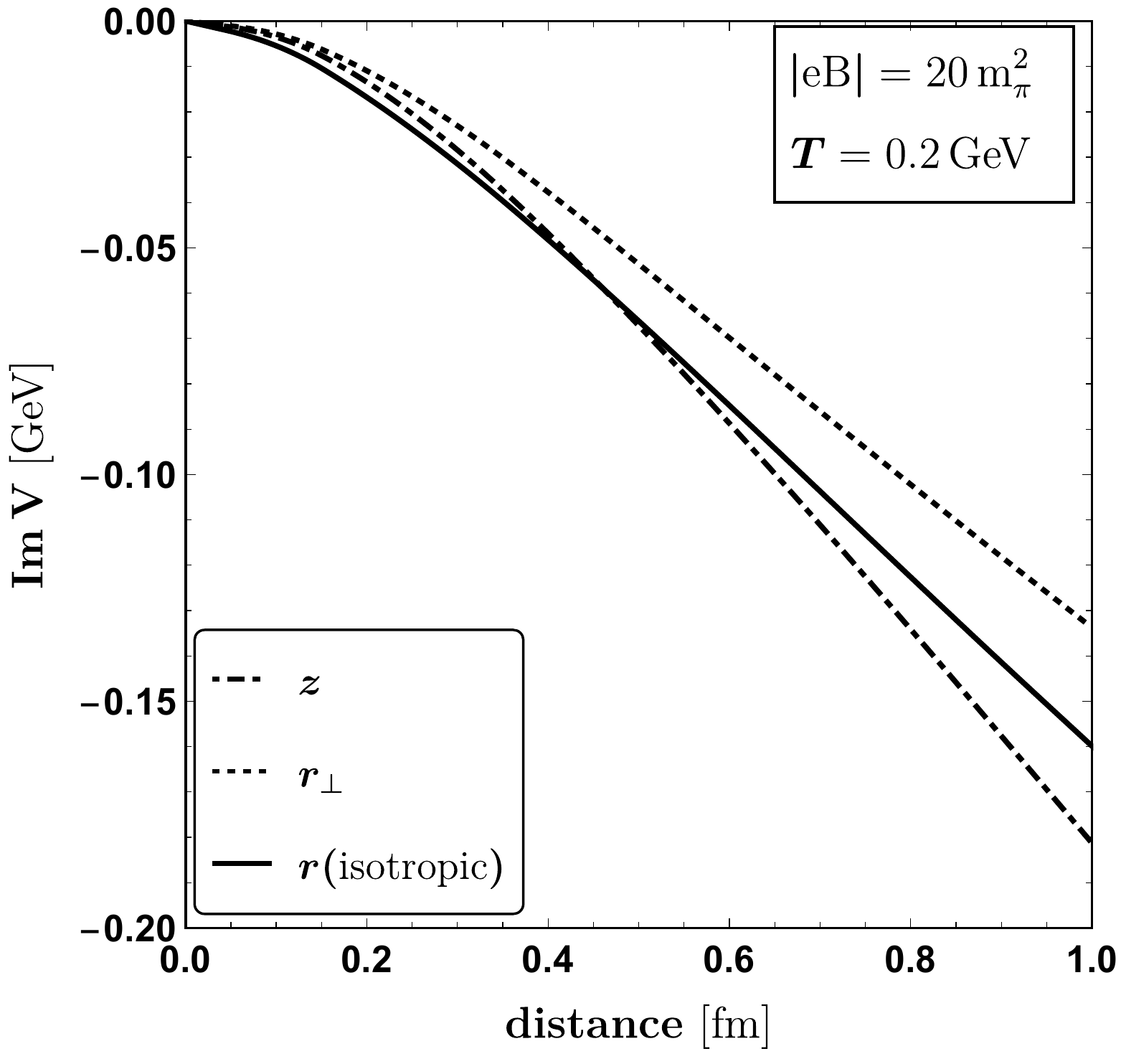}
\caption{Variation of ${\rm Im}~V$ with distance comparing between full and Debye mass approximated expression. The variation is shown with $z$ for vanishing $r_\perp$ (dashed curve), with $r_\perp$ for vanishing $z$ (dotted curve), and with isotropic $r$ using Debye mass approximated expression (solid curve). }
\label{structure}
\end{center}
\end{figure}

In fig.~\ref{structure}, we have again compared our general result with the Debye mass approximated results, but this time for arbitrary values of external fields. As our observable, again we have chosen the imaginary part of the HQ potential. In appendix~\ref{appD}, we have given the expression for the Debye mass approximated imaginary part of the HQ potential, which is isotropic in nature (i.e. no explicit dependence on $r_\perp$ and $z$), unlike our most general result. So, in fig.~\ref{structure}, we have shown this Debye mass approximated isotropic curve (solid curve) with our anisotropic curves (dashed and dotted curves). The anisotropic curves are plotted from our main result, i.e. eq.~\eqref{imag_vr_final} using eq.~\eqref{imD00}. Among the two curves, the dashed curve shows the variation with $z$ for vanishing $r_\perp$ whereas the dotted curve shows the variation with $r_\perp$ for vanishing $z$. On the other hand, the solid curve is drawn using eq.~\ref{imvgen} where one can see that the magnetic field effect is coming solely through the Debye mass, subsequently providing incomplete information. In this scenario, we are getting the isotropic space dependency of the potential. One can observe from fig.~\ref{structure} that the difference between the full result and the Debye mass approximated result increases significantly with increasing distance, thereby emphasizing the importance of considering the anisotropic nature of the full HQ potential in presence of arbitrary values of external magnetic fields. So, from Fig.~(\ref{imagV_comparison}) to (\ref{structure}) are devoted to show our ingredient details in the heavy quark potential at finite $T$, $B$ with respect to earlier calculations~\cite{Singh:2017nfa,Hasan:2020iwa}. Next, we will zoom-in more the anisotropic tomography of this heavy quark potential due to magnetic field, which is probably first time addressed in the literature.    

\begin{figure} 
\begin{center}
\includegraphics[scale=0.5]{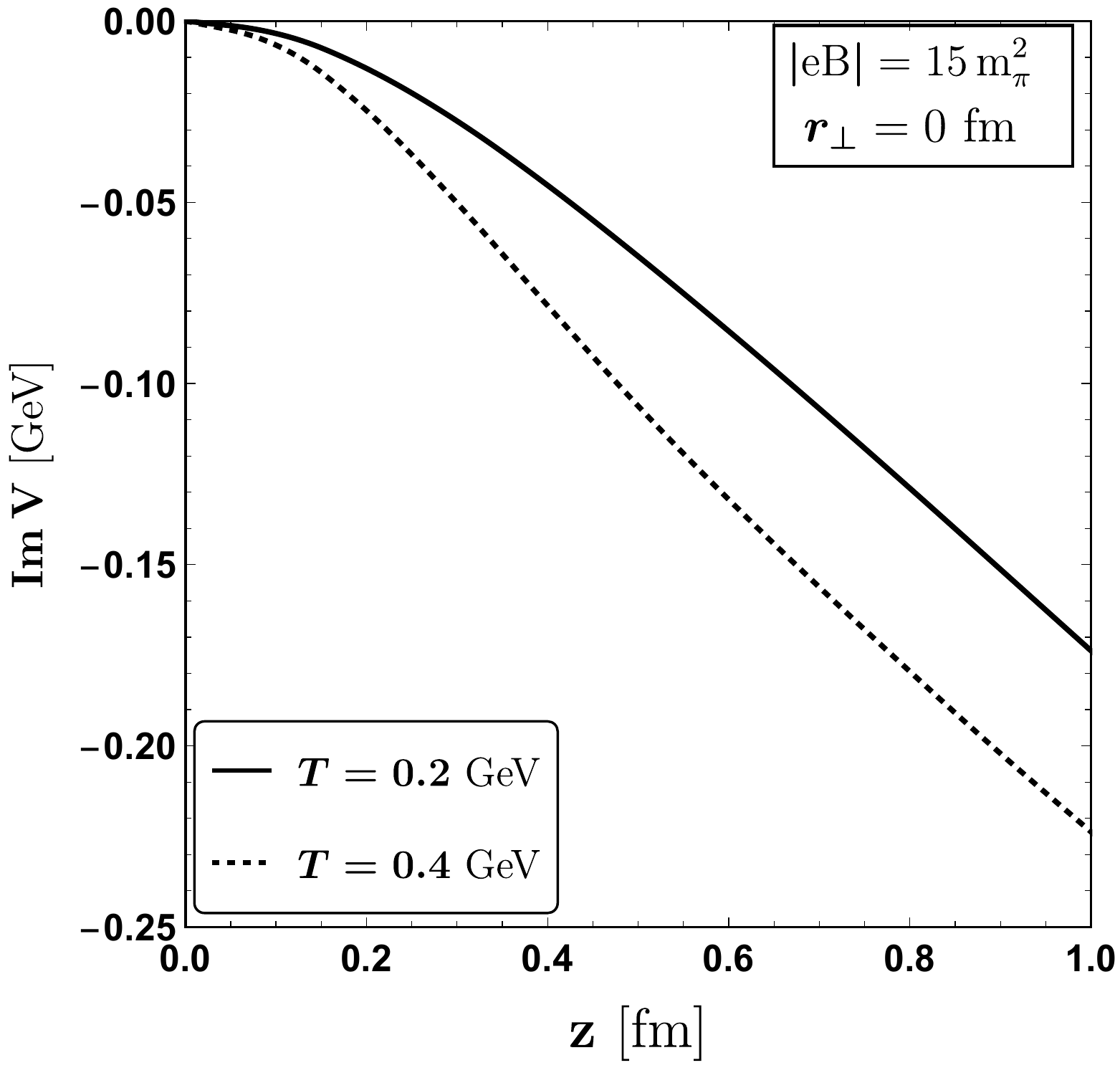}
\includegraphics[scale=0.5]{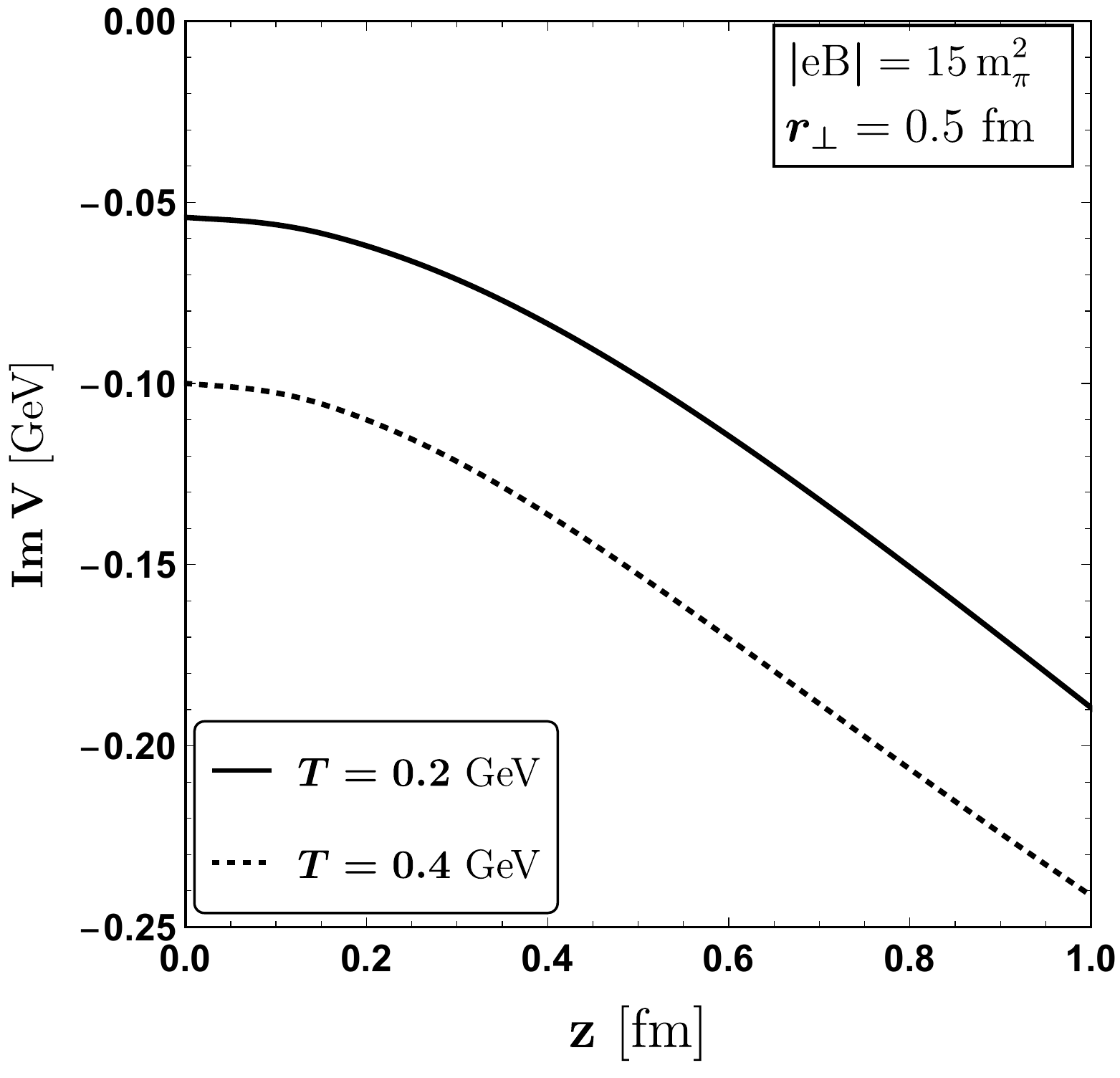}
\includegraphics[scale=0.5]{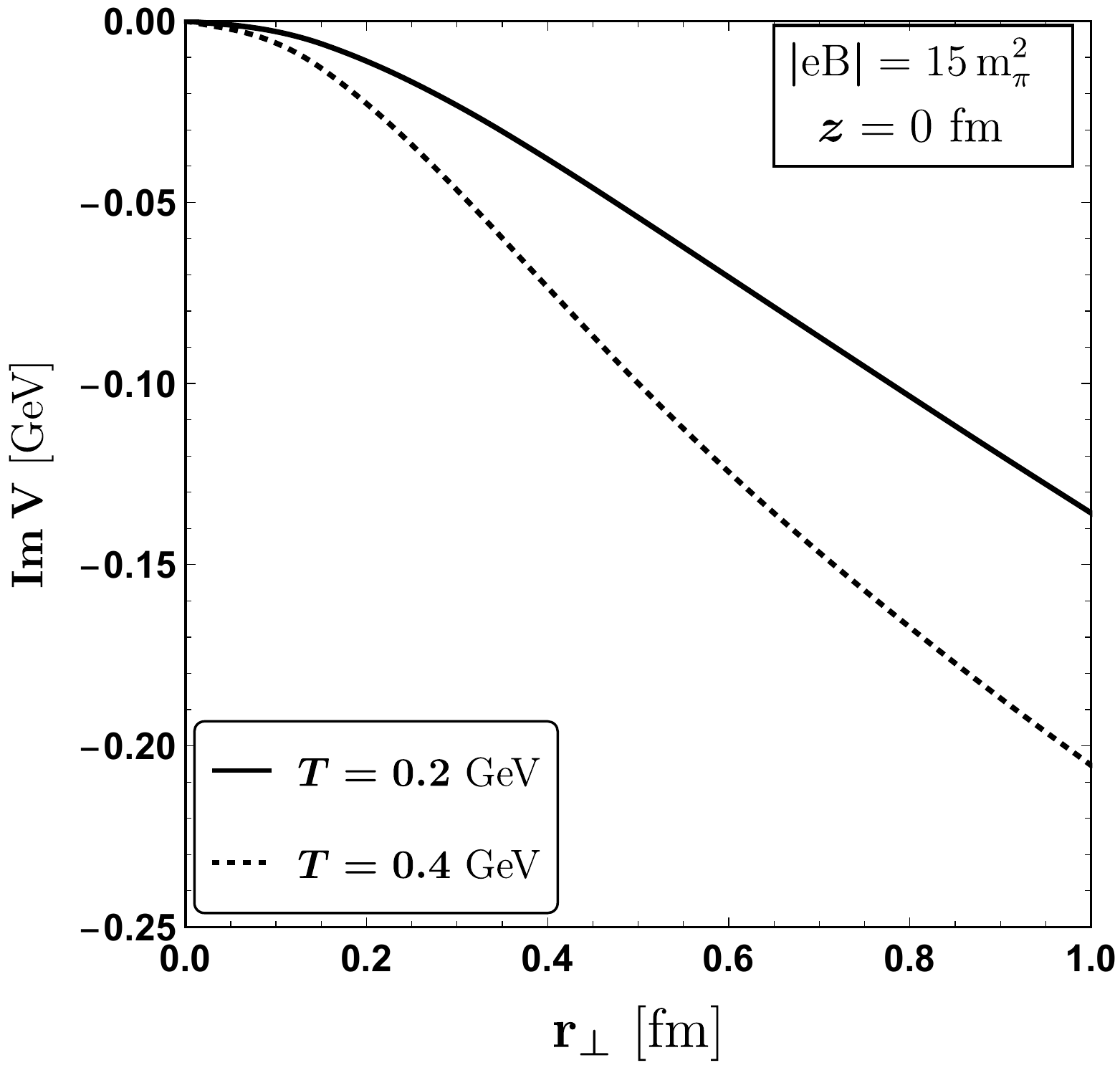}
\includegraphics[scale=0.5]{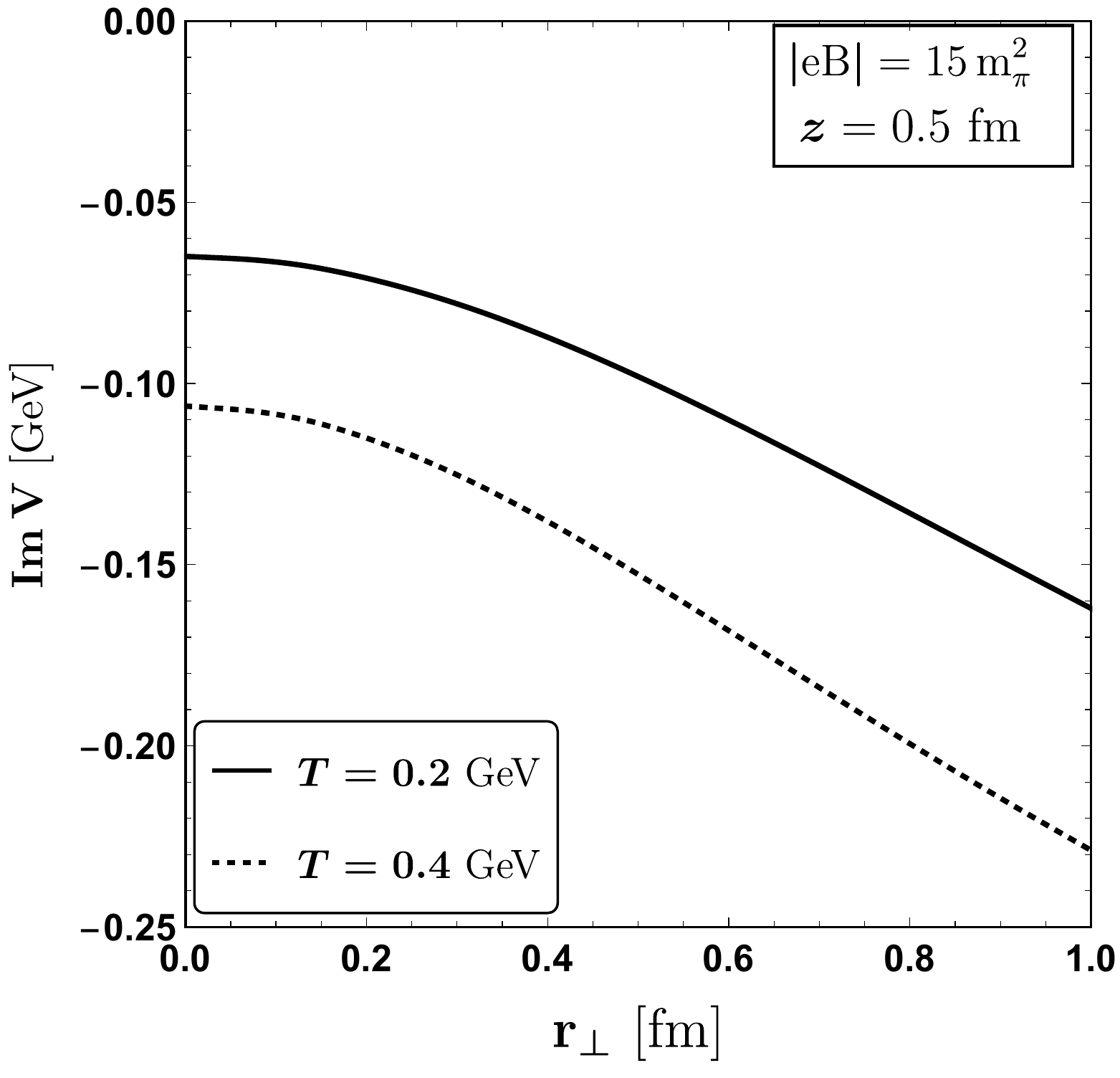}
\caption{Variation of ${\rm Im}~V$ with $z$ for two different fixed values of $r_\perp$ - $r_\perp = 0$ (upper-left panel) and $r_\perp =0.5$ fm (upper-right panel) and with $r_\perp$ for two different fixed values of $z$ - $z = 0$ (lower-left panel) and $z =0.5$ fm (lower-right panel). For each of the plots we have chosen two different values of temperature and a fixed value of the external magnetic field.}
\label{imagV_aniso1}
\end{center}
\end{figure}

\begin{figure}[tbh]
\begin{center}
\includegraphics[scale=0.5]{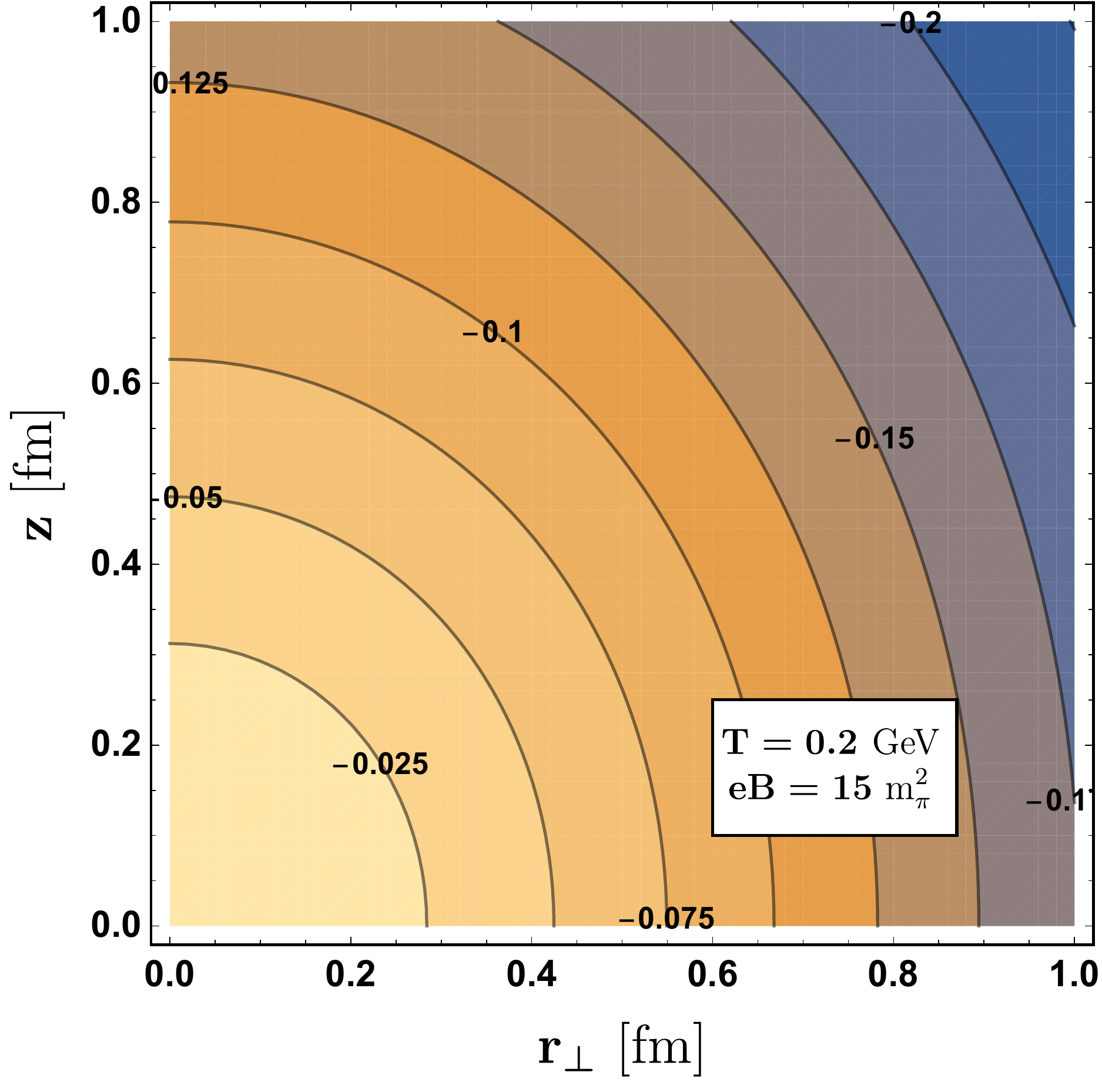}
\caption{Contour plot of ${\rm Im}~V$ showing the equal potential regions for different values of $r_\perp$ and $z$.}
\label{imagV_contour}
\end{center}
\end{figure}

Fig.~\ref{imagV_aniso1} have explored the anisotropic nature of the heavy quark potential in presence of the external magnetic field, applied along the $z$ direction. In upper panel of fig.~\ref{imagV_aniso1} we have shown the variation of the imaginary part of the HQ potential with respect to the longitudinal distance $z$ for two different fixed values of the transverse distance $r_\perp = 0$ (upper-left panel) and $r_\perp = 0.5$ fm (upper-right panel). For both the plots we have fixed the external magnetic field to $eB = 15 m_\pi^2$ and shown the variation for two different values of the temperature, i.e. $T=0.2$ and $T=0.4$ GeV. For the plot with $r_\perp = 0$, at lower values of $z$, both the curves start from vanishing ${\rm Im}~V$, as expected. Also for both the plots one can notice that with higher values of temperature, the magnitude of ${\rm Im}~V$ also becomes higher. The curves show a gradually decreasing behaviour of the imaginary part of the HQ potential with increasing distance, as was also evident from figures.~\ref{imagV_comparison} and \ref{imagV_LLLvsfull}. Lower panel of Fig.~\ref{imagV_aniso1} shows similar behaviours, where we have fixed $z$ to two different values of $z=0$ (lower-left panel) and $z=0.5$ fm (lower-right panel) and varied ${\rm Im}~V$ with respect to $r_\perp$. 
\begin{figure}[tbh]
\begin{center}
\includegraphics[scale=0.5]{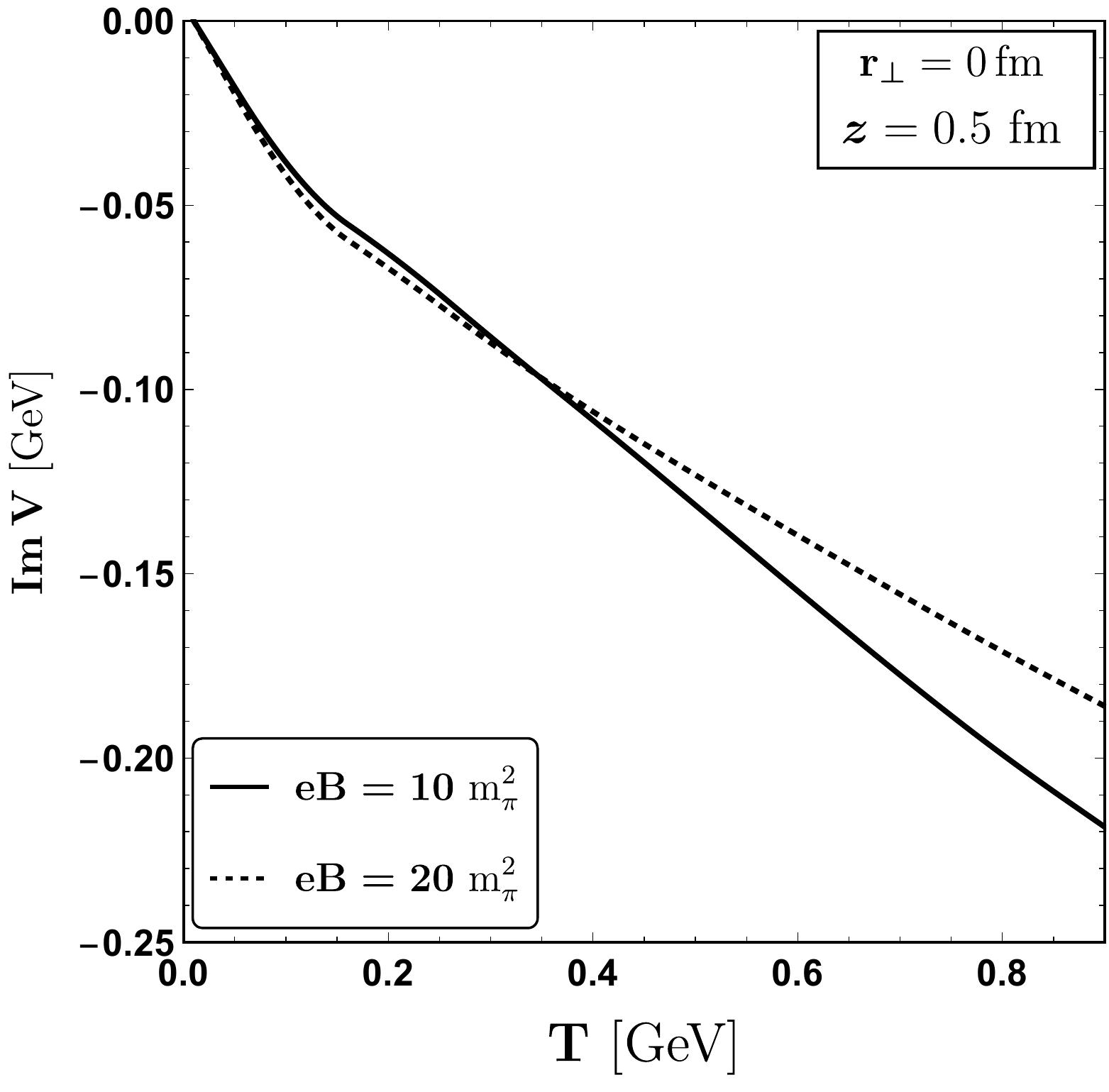}
\includegraphics[scale=0.5]{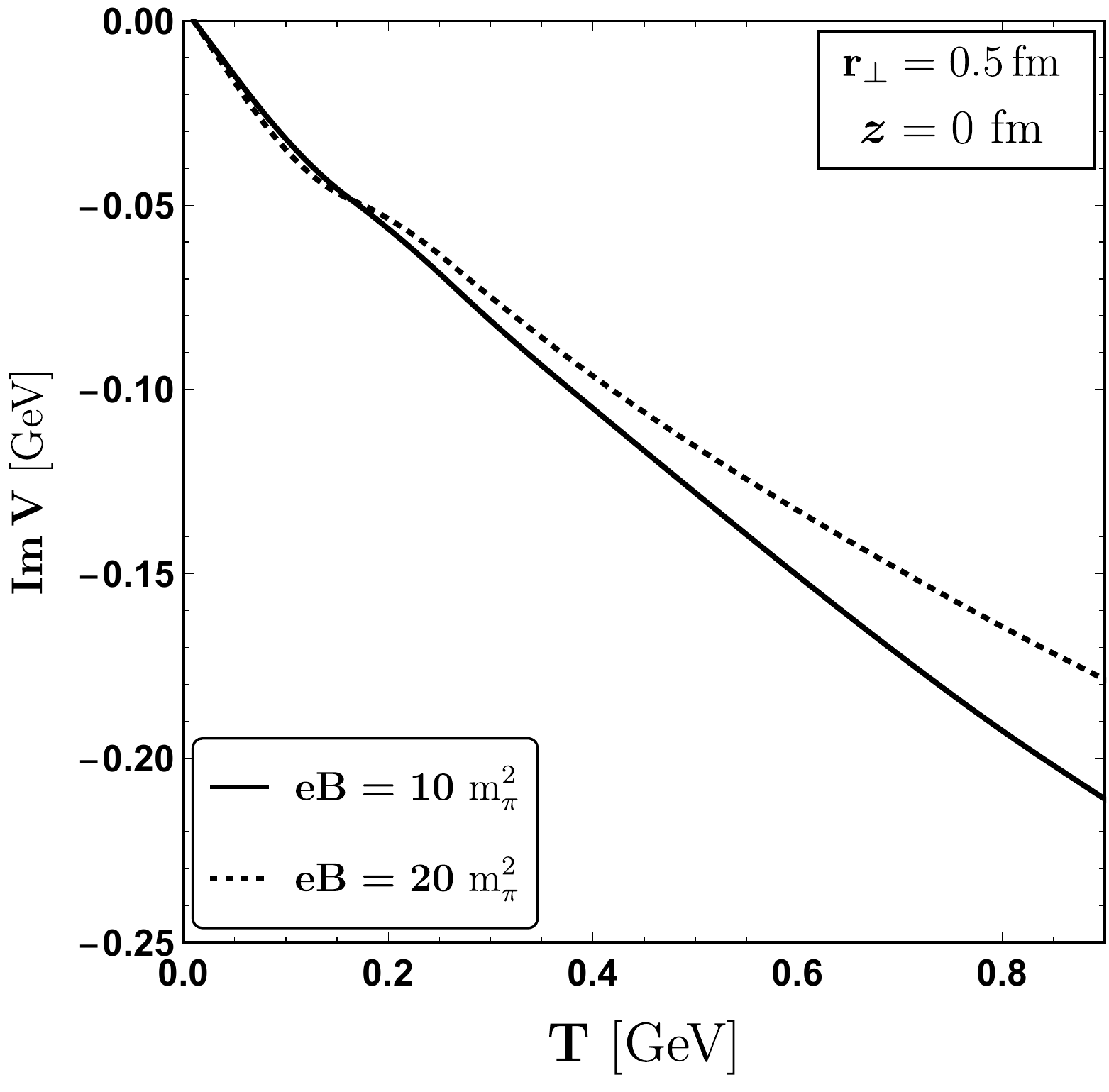}
\caption{Variation of ${\rm Im}~V$ with temperature shown for two different cases, i.e. for vanishing $r_\perp$ (left panel) and for vanishing $z$ (right panel). for each of the plots, we have chosen two different values of the external magnetic field which shows some interesting crossovers.}
\label{imagV_varyT}
\end{center}
\end{figure}

In fig.~\ref{imagV_contour} we have shown the overall spatial dependence of the imaginary part of the HQ potential in the form of a contour plot where we have varied both $z$ and $r_\perp$ within the range of 0 to 1 fm. For this plot, we have fixed the values of the temperature and the magnetic field as 0.2 GeV and 15 $m_\pi^2$ respectively. The equal potential (imaginary) regions are represented by different curves and the corresponding values for the imaginary parts of the HQ potential is depicted on top of each curve. Reader should notice that equi-potential curves are elliptic in nature instead of circle. Finite magnetic field make this transformation from circle to ellipse, meaning isotropic to anisotropic transformation. So this anisotropic tomography of heavy quark potential may be used as a signature of magnetic field, produced in heavy ion collision. Though this task is very non-trivial but we will try to search the possibility by presenting our results in different angle. The contour plot (fig.~\ref{imagV_contour}) will be modified with temperature and magnetic field, which is explored in next paragraph.

In fig.~\ref{imagV_varyT}, we have presented the variation of the imaginary part of the HQ potential with respect to the temperature for two different values of the external magnetic field, i.e. $eB=10 m_\pi^2$ and $eB = 20 m_\pi^2$. We have considered the case of vanishing transverse distance in the left panel with a fixed value of $z=0.5$ fm. It can be observed that for vanishing temperatures, curves for different magnetic field merges into giving a vanishing ${\rm Im}~V$ as in-medium dissociation phenomena of quarkonia can't be expected in vacuum or $T=0$. When the temperature starts to increase gradually, at first the curve for the higher magnetic field gives higher values for imaginary part of the HQ potential. However, after a certain temperature, we observe a crossing between the curves. This feature can be understood in the following way: In the low temperature region, magnetic field is the most dominating scale. As the temperature starts to increase, a competition between the magnetic field and the temperature takes place. The nature of the curves get inverted with the enhancement of the temperature, as the temperature scale becomes more dominant for $eB = 10 ~m_\pi^2$, compare to $eB = 20 ~m_\pi^2$.
Similar behaviors have also been observed in the right panel where we have vanishing $z$ and a fixed $r_\perp = 0.5$ fm. 
So, according to Fig.~\ref{imagV_varyT}, dissociation probability is enhanced and suppressed by magnetic field in the low and high temperature respectively. If we concentrate within $T=0.1$-$0.4$ GeV, $eB=10$-$20 m_\pi^2$ as covering domain of expanding QGP and heavy quark dissociation temperature range broadly as $T_d=0.15$-$0.35$ GeV, then along $z$-axis, dissociation probability can be enhanced due to magnetic field, while opposite impact of magnetic field can be occurred along $r_\perp$-axis. This comment is based on the left and right panels of Fig.~\ref{imagV_varyT}, which are plotted at $r_\perp=0$, $z=0.5$ fm and $r_\perp=0.5$ fm, $z=0$ respectively. However, for exact knowledge of enhancing and suppressing dissociation domain,
one should notice the variation of all four parameters - $r_\perp$, $z$, $T$ and $eB$.
%

From the earlier discussion, we can see a rich anisotropic tomography of heavy quark dissociation by varying $r_\perp$, $z$, $T$ and $eB$ but
when we go towards experimental quantity - heavy quark dissociation probability, this anisotropic tomography will be integrated and we will get only temperature and magnetic field dependent dissociation, which carry the anisotropic information through its integrated values, which will be different from corresponding integrated values isotropic potential. Using Eq.~(\ref{Gamma}), this integrated values of heavy quark dissociation are obtained. 
%
\begin{figure} 
\begin{center}
\includegraphics[scale=0.5]{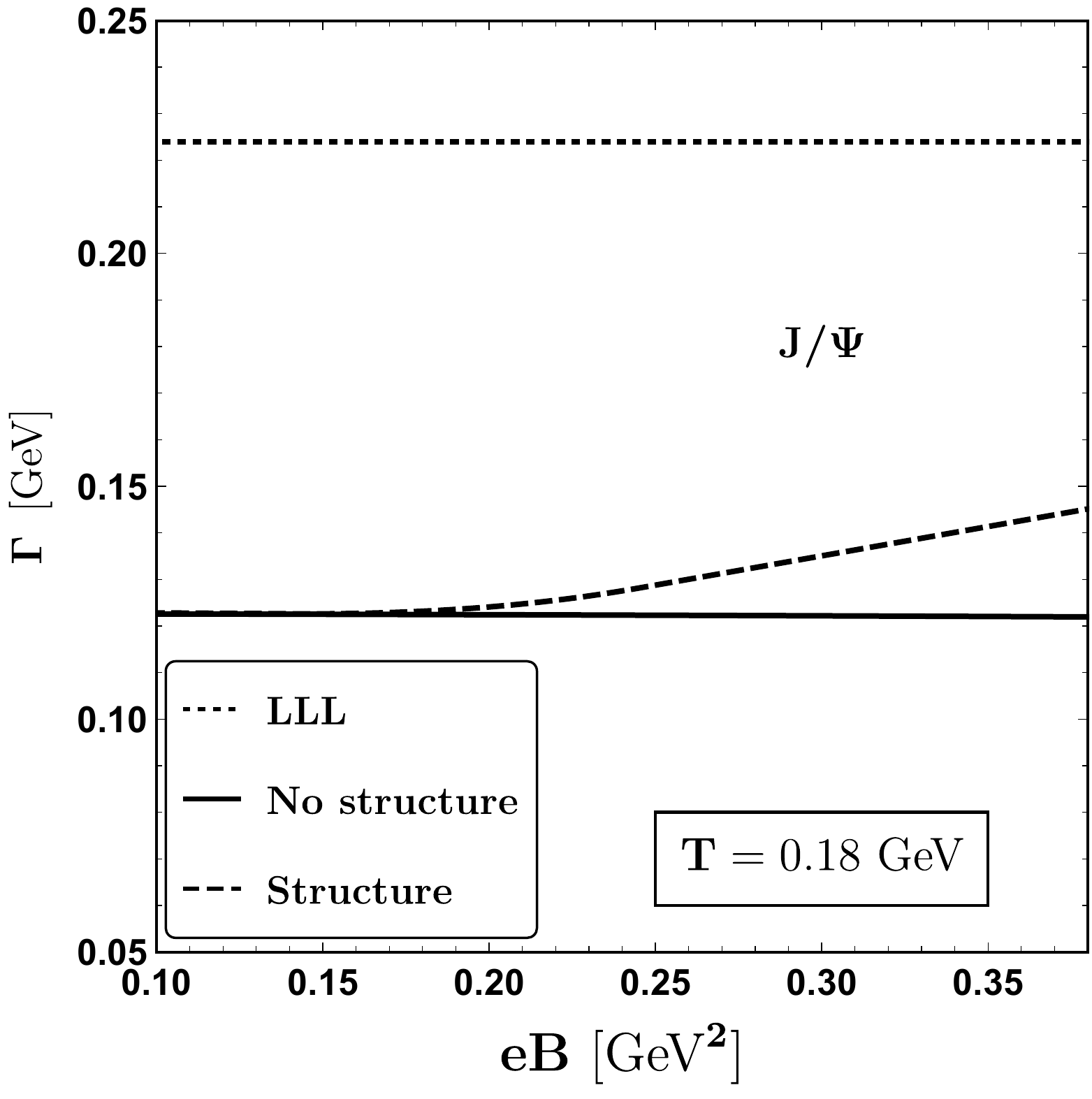}
\includegraphics[scale=0.5]{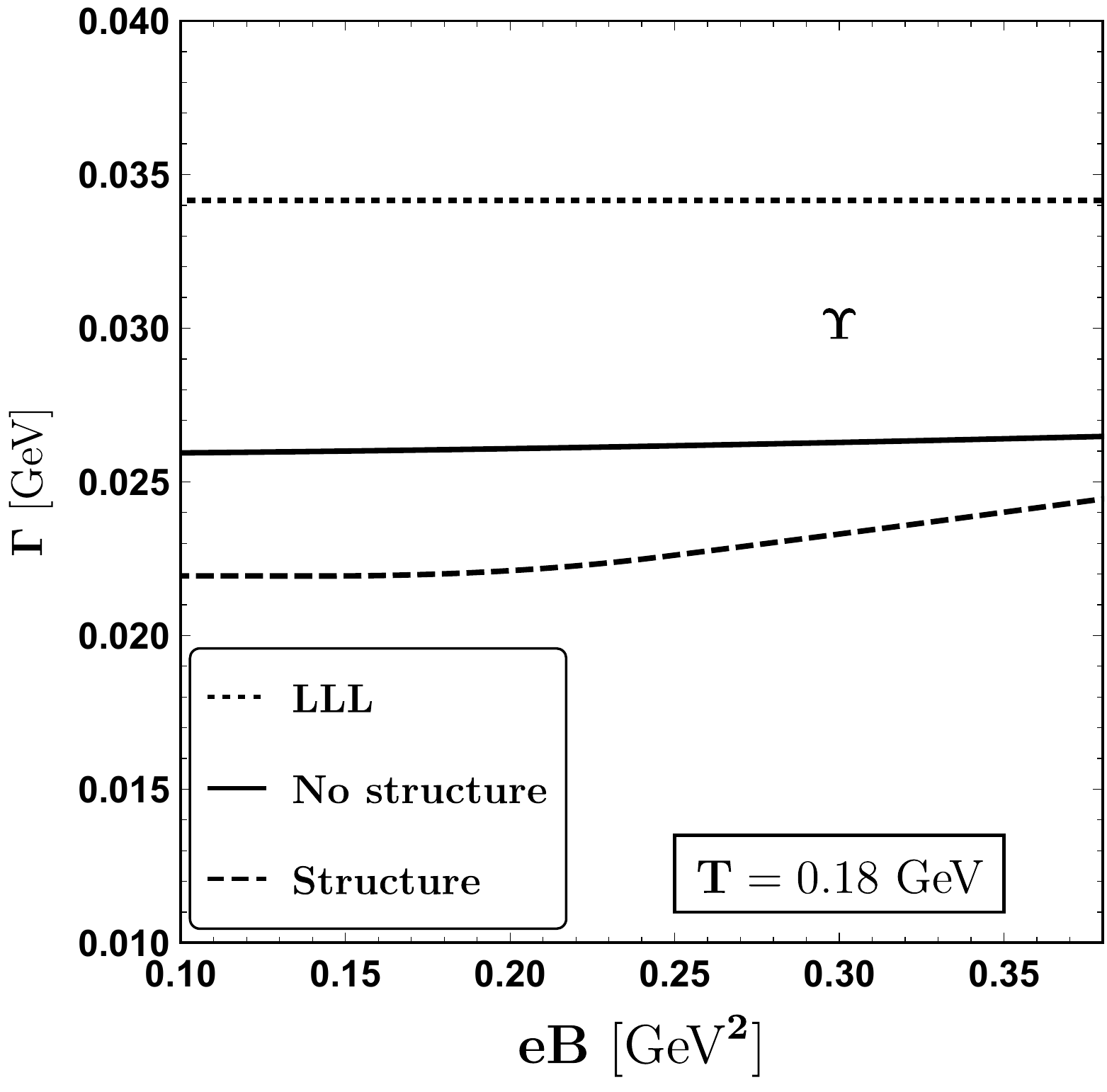}
\caption{Variation of the decay width $\Gamma$ with respect to the external magnetic field for a fixed temperature shown for the case of Charm quark (left panel) and Bottom quark (right panel). Curves shown for the case of LLL approximated result, $m_D$ approximated result and full result.}
\label{Gamma_varyeB}
\end{center}
\end{figure}
\begin{figure}[tbh]
\begin{center}
\includegraphics[scale=0.5]{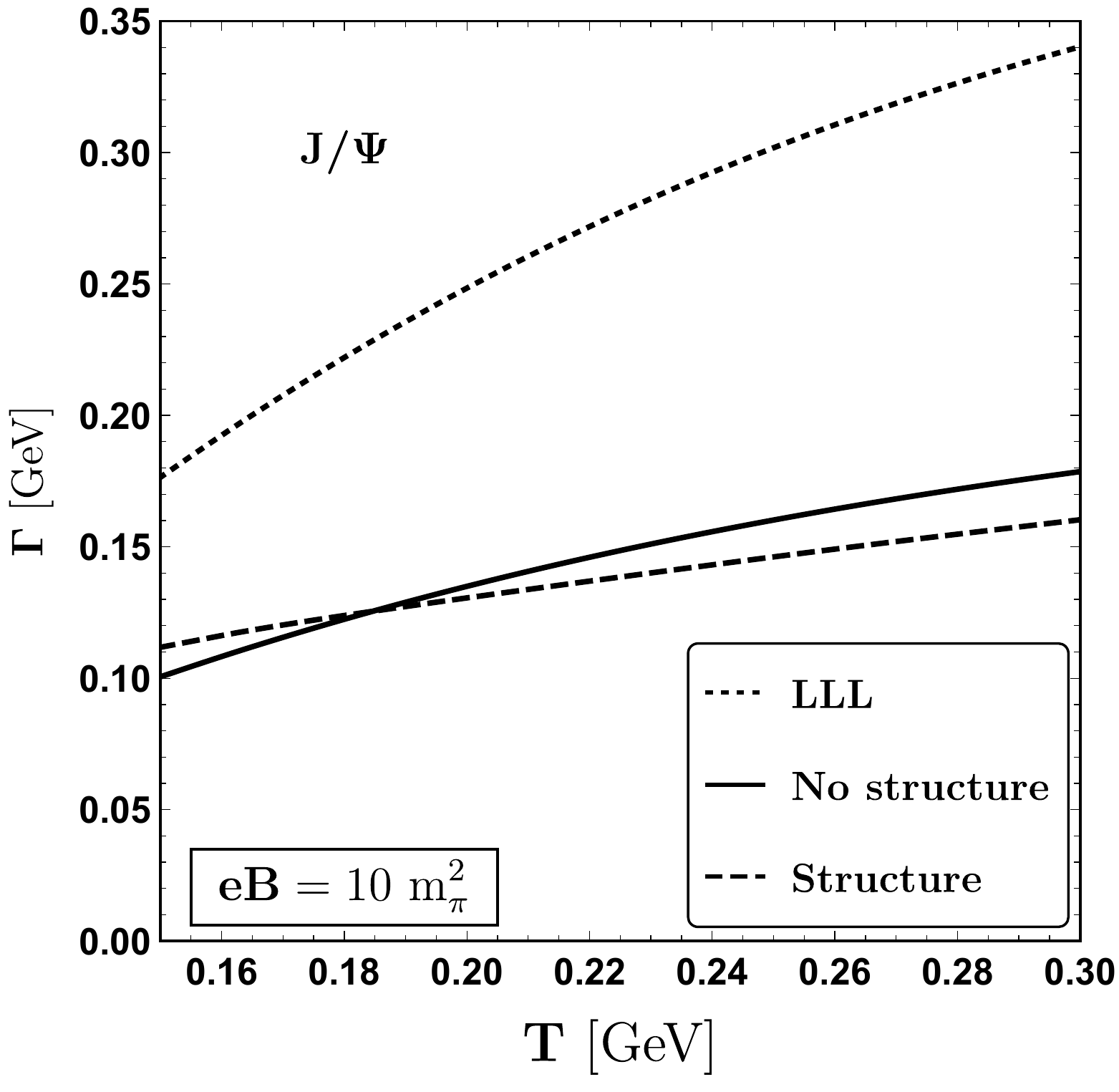}
\includegraphics[scale=0.5]{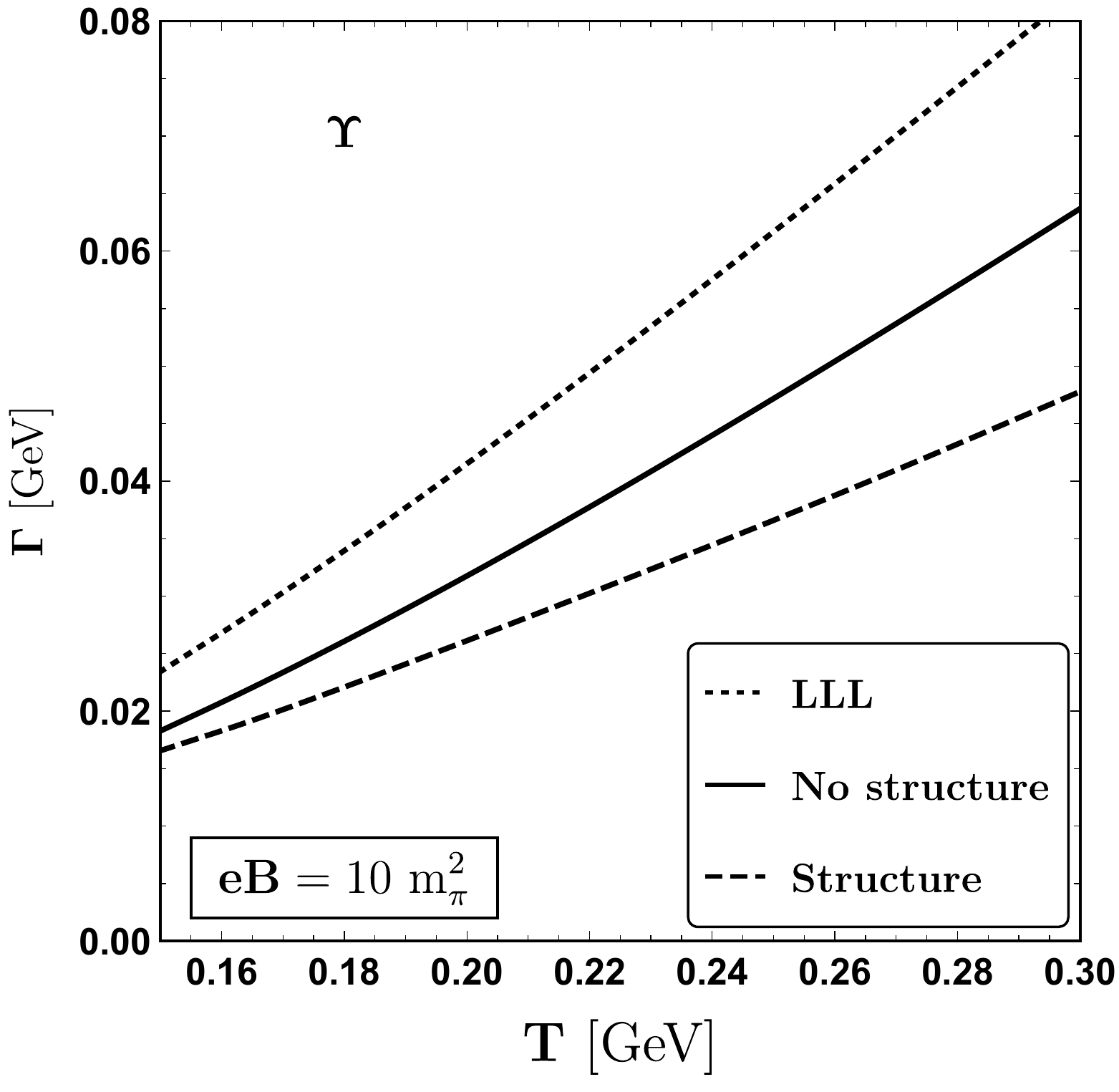}
\caption{Variation of the decay width $\Gamma$ with respect to the temperature for a fixed external magnetic field shown for the case of Charm quark (left panel) and Bottom quark (right panel). Curves shown for the case of LLL approximated result, $m_D$ approximated result and full result.}
\label{Gamma_varyT}
\end{center}
\end{figure}
In figs.~\ref{Gamma_varyeB} and \ref{Gamma_varyT}, we have studied the variation of the decay width with respect to the external magnetic field and temperature respectively. In the calculation we take the bottomonium and charmonium masses as $m_b=4.66\, \text{GeV}$ and  $m_c=1.275\, \text{GeV}$ respectively~\cite{ParticleDataGroup:2014cgo,Singh:2017nfa}. For each of the cases, we have shown two plots, for charm (left panel) and bottom (right panel) quarks. In fig.~\ref{Gamma_varyeB} we have fixed the temperature at $T=180$ MeV and in fig.~\ref{Gamma_varyT} we have fixed the magnetic field at $eB = 10 m_\pi^2$. In each of the plots we have compared our full result (dashed lines) with the LLL approximated result (dotted lines) and the Debye mass approximated result without structure (solid lines). One can notice from fig.~\ref{Gamma_varyeB} that the LLL approximation again overestimates the magnitude of the decay width. In comparison with the Debye mass approximated result, our full result of the decay width shows 
a different $T$ and $eB$ profile
for both charm and bottom quarks. 
An increasing behaviour with increasing temperature can also be found in fig.~\ref{Gamma_varyT}.
As the bottomonium states are smaller in size with larger masses than the charmonium states, the thermal width for $\Upsilon$ is considerably smaller than the $J/\Psi$. 

At the end, we want to emphasise once again that with respect to earlier estimations~\cite{Hasan:2020iwa,Singh:2017nfa} of heavy quark dissociation in presence of finite magnetic field, present results find a new dimension, i.e. a new profile in temperature and magnetic field axes and more intriguingly a rather complex anisotropic tomography in heavy quark dissociation. Former modification is found for adopting the general structure of gluon propagator at finite magnetic field in heavy quark potential framework, which is done here for first time. On the other hand, the later modification - anisotropic tomography of heavy quark dissociation, which can always be expected at finite magnetic field if one carefully considered parallel and perpendicular momentum components during Fourier's transformation. Since earlier Refs.~\cite{Hasan:2020iwa,Singh:2017nfa} have not included this consideration, so anisotropy structure is missing in their calculations.
%
In this context, our present work is first time pointing out this anisotropic structure in heavy quark potential, which might build an anisotropic dissociation. In near future, our plan is to connect this anisotropic aspect of heavy quark potential due to magnetic field with quarkonia suppression phenomenology, which might unfold how to get signature strong magnetic field through quarkonia phenomenology.

\section{Conclusions}
\label{sec10}
In the present theoretical study, we have evaluated the imaginary part of heavy quark complex potential formalism at finite temperature and magnetic field, whose preliminary steps are standard and as follows. The imaginary part of heavy quark-antiquark potential in terms of coordinate space, temperature and magnetic field can be estimated by taking Fourier's transform of momentum dependent potential, divided by permittivity of the medium, which carry temperature and magnetic field. This permittivity can be calculated from the temporal component of the effective gluon propagator at finite temperature and magnetic field. 

Now, in the present work we have adopted the general structure of the gluon propagator at finite temperature and magnetic field, which was not considered in earlier works. So a new ingredient of temperature and magnetic field field dependent profile in calculations is found. Our adopted generalized gluon propagator consisted four linearly independent tensors. So there are four form factors which can be calculated from the gluon self-energy. In our case, we have only needed one form factor explicitly for the sake of our calculation. This is evaluated from the one loop gluon self-energy where the quark loop is affected by magnetic field. So the modified quark propagator in presence of magnetic field is considered. We have obtained the results for general magnetic field by summing all Landau level contributions. Comparing our results with the existing works, done with lowest Landau level approximation in the strong field limit as well as weak field approximation, one can consider our work as the more general in nature. Our results are applicable for the entire range from weak to strong magnetic field. This is because we are considering all Landau level summation and most general structure of gluon propagators, which are taking care of corresponding full quantum mechanical and quantum field theoretical effects respectively.
	
Apart from these new ingredients - all Landau level summation and general structure of propagator, present work has adopted another novel and interesting fact - anisotropic form of heavy quark potential in presence of magnetic field, which were ignored in earlier works, because of some approximations like ignoring the Debye mass independent terms. We have graphically presented the detailed anisotropic tomography of imaginary part of potential, which modifies with temperature and magnetic field. This is one of the main findings of our present study, which to the best of our knowledge, has not been discussed before in the literature for heavy quark potential.  
	
%

Imaginary part of heavy quark potential basically provides us with its dissociation probability. After doing the co-ordinate space integration by folding with probability density, based on simple wave-function due to Coulomb-type potential, we have obtained the temperature and magnetic field dependent dissociation probability or thermal width of quarkonium states - $J/\Psi$ and $\Upsilon$. Here, we have again found the modified temperature and magnetic field profile due to considering the summation over all possible Landau levels and the general structure of gluon propagator at finite magnetic field with respect to earlier references. We believe that the anisotropic aspect of heavy quark potential due to magnetic field might build an interesting quarkonia phenomenology, which is planned for our next work. Further studies on the angular dependence / ellipticity (e.g. for the case of photon emission, see ~\cite{Wang:2020dsr} ) of the dissociation probability is also needed to disentangle the anisotropy due to the external magnetic field with the geometrical effects coming from the shape of the plasma produced in non central HIC.


\section*{Acknowledgement}
RG, AB and IN acknowledges IIT Bhilai for the academic visit and hospitality during the course of this work. They thank Shivani Valecha, Sunima Baral, Srikanta Debata, Purushottam Sahu, Naba Kumar Rana for helping them in arranging accommodation. RG is funded by University Grants Commission (UGC). AB is supported by the Guangdong Major Project of Basic and Applied Basic Research No. 2020B0301030008, Science and Technology Program of Guangzhou Project No. 2019050001 and the postdoctoral research fellowship from the Alexander von Humboldt Foundation, Germany. IN acknowledges the Women Scientist Scheme A (WoS A) of the Department of Science and Technology (DST) for the funding with grant no. DST/WoS-A/PM-79/2021.

\appendix

\section{Gluon effective propagator in presence of magnetic field}
\label{appA}

In presence of thermal medium, the Lorentz (boost) invariance is broken, whereas the presence of magnetic field breaks the rotational symmetry of the system. Heat bath velocity $u^\mu=(1,0,0,0)$ is introduced in presence of thermal medium. We consider the magnetic field along $z$ direction {\it{i.e.,}} $n_\mu=(0,0,0,1)$. We define $\bar n^\mu=A^{\mn} n_\mu$. Now gluon self energy in the presence of thermomagnetic medium can be written as
	\bea
	\Pi^{\mn}=b B^{\mn}+c R^{\mn}+dQ^{\mn}+a N^{\mn}\label{self_energy}
	\eea
	where  the basis tensors are given as~\cite{Karmakar:2018aig}
	\bea
	B^{\mn}&=&\frac{\bar u^\mu \bar u^\nu}{\bar u^2},\\
	Q^{\mn}&=&\frac{\bar n^\mu \bar n^\nu}{\bar n^2},\\
	N^{\mn}&=&\frac{\bar u^\mu \bar n^\nu+\bar u^\nu \bar n^\mu}{\sqrt{\bar u^2}\sqrt{\bar n^2}},\nn\\
	R^{\mn}&=&V^{\mn} -B^{\mn}-Q^{\mn}.
	\eea
	$b$, $c$, $d$ and $a$ are the corresponding form factors. The vacuum projection tensor is 
	\bea
	V^{\mn}=g^{\mn}-\frac{P^\mu P^\nu}{P^2}.
	\eea
	$\bar u^\mu$ is defined by projecting the vacuum projection tensor upon $u^{\mu}$ i.e. $\bar u^\mu=V^{\mn}u_\nu$ and $\bar n^\mu$ is defined as $\bar n^\mu=A^{\mn}n_\nu$.
	The form factors can be calculated using the following relations.
	\bea
	b&=&b_g+b_q=-\frac{p_0^2-p^2}{p^2}\bigg[\Pi^{g}_{00}(P) +\Pi^{q}_{00}(P)\bigg] \label{b},\\
	c&=&c_g+c_q=R^{\mu\nu}\Big[\Pi_{\mn}^g(P)+\Pi^{q}_{\mu\nu}(P)\Big]\nn\\
	&=&  (\Pi^{g})^{\mu}_\mu(P) +(\Pi^{q})^{\mu}_\mu(P) + \frac{1}{p_\perp^2}\Big[\(p_0^2-p_\perp^2\)\Big\{\Pi_{00}^{g}(P)+\Pi_{00}^{q}(P)\Big\} p^2\Big\{\Pi_{33}^{g}(P)+\Pi_{33}^{q}(P)\Big\}\nn\\
	&&-2p_0p_3\Big\{\Pi_{03}^{g}(P)+\Pi_{03}^{q}(P)\Big\}\Big],
	\label{c}\\
	d&=&d_g+d_q=Q^{\mu\nu}\Big[\Pi_{\mn}^g(P)+\Pi^{q}_{\mu\nu}(P)\Big]\nn\\
	&=&-\frac{p^2}{p_\perp^2}
	\bigg[\Big\{\Pi_{33}^{g}(P)+\Pi_{33}^{q}(P)\Big\} -\frac{2p_0p_3}{p^2}\Big\{\Pi_{03}^{g}(P)+\Pi_{03}^{q}(P)\Big\} + \frac{p_0^2p_3^2}{p^4}\Big\{\Pi_{00}^{g}(P)+\Pi_{00}^{q}(P)\Big\}\bigg], \label{d}\\
	a&=&a_g+a_q=\frac{1}{2}N^{\mn}\Big[\Pi_{\mu\nu}^{g}+\Pi_{\mu\nu}^{q}\Big]\nn\\
	&=&\frac{1}{2\sqrt{\bar u^2}\sqrt{\bar n^2}}\bigg[-2\frac{\bar u \cdot n}{\bar u^2}\Big\{\Pi_{00}^{g}+\Pi_{00}^{q}\Big\}+2\Big\{\Pi_{03}^{g}+\Pi_{03}^{q}\Big\}\bigg]\label{a}.
	\eea
	where $\Pi_{\mn}^g$ and $\Pi_{\mn}^q$ are the self energy contributions from the gluon loop, ghost loop and from the quark loop respectively. The form factors would be calculated from one loop gluon self energy diagram. 
	
	The general structure of the gluon effective propagator using Eq.~\eqref{self_energy} is given as~\cite{Karmakar:2018aig}
	\bea
	D^{\mn}&=& \frac{\xi P^\mu P^\nu}{P^4}+\frac{P^2-d}{(P^2-b)(P^2-d)-a^2}B^{\mn}+\frac{1}{P^2-c}R^{\mn}+\frac{P^2-d}{(P^2-b)(P^2-b)-a^2}Q^{\mn}\nn\\
	&+&\frac{a}{(P^2-b)(P^2-d)-a^2}N^{\mn}.
	\label{eff_prop}
	\eea

\section{Frequency sum}
We write the fermionic Matsubara sums. Here $ \om_n=(2n+1)\pi T$ and $\om_m=2m\pi T $ are the fermionic and bosonic Matsubara frequencies respectively.
\bea
&&T\sum_{n=-\infty}^{\infty} \frac{1}{[(i\om_n)^2-E_k^2][(i\om_n-i\om_m)^2-E_q^2]}\nn\\
&=&\sum_{s_1,s_2=\pm 1}\frac{1}{4s_1E_k E_q}\frac{n_F(E_{k})-n_F(s_1 E_{q})}{is_2 \om_m+E_{k}-s_1E_{q}},
\eea
and
\bea
&&T\sum_{n=-\infty}^{\infty} \frac{i\om_n (i\om_n-i\om_m)}{[(i\om_n)^2-E_k^2][(i\om_n-i\om_m)^2-E_q^2]}\nn\\
&=&\sum_{s_1,s_2=\pm 1}\frac{1}{4}\frac{n_F(E_{k})-n_F(s_1 E_{q})}{is_2 \om_m+E_{k}-s_1E_{q}}.
\eea
The fermi-Dirac distribution function is given as, $n_F(E)=\frac{1}{\exp(E/T)+1}$.

\section{Definition of functions $X_{m,n}$ and $X^1_{m,n}$}
\label{app:B}
\bea
X_{m,n}&=& \frac{m!}{n!} e^{-p_\perp^2 d_f^2/2}\bigg(\frac{p_\perp^2 d_f^2}{2}\bigg)^{n-m} \bigg[L_m^{n-m}\bigg(\frac{p_\perp^2 d_f^2}{2}\bigg)\bigg]^2 ,\,\,\,\, \text{for}\,\, n\geq m\\
&=& \frac{n!}{m!} e^{-p_\perp^2 d_f^2/2}\bigg(\frac{p_\perp^2 d_f^2}{2}\bigg)^{m-n} \bigg[L_n^{m-n}\bigg(\frac{p_\perp^2 d_f^2}{2}\bigg)\bigg]^2 ,\,\,\,\, \text{for}\,\, n < m\\
X_{m,n}^1&=& 2\frac{(m+1)!}{n!} e^{-p_\perp^2 d_f^2/2}\bigg(\frac{p_\perp^2 d_f^2}{2}\bigg)^{n-m} L_m^{n-m}\bigg(\frac{p_\perp^2 d_f^2}{2}\bigg) L_{m+1}^{n-m}\bigg(\frac{p_\perp^2 d_f^2}{2}\bigg),\,\,\,\, \text{for}\,\, n\geq m\\
&=&2 \frac{(n+1)!}{m!} e^{-p_\perp^2 d_f^2/2}\bigg(\frac{p_\perp^2 d_f^2}{2}\bigg)^{m-n} L_n^{m-n}\bigg(\frac{p_\perp^2 d_f^2}{2}\bigg) L_{n+1}^{m-n}\bigg(\frac{p_\perp^2 d_f^2}{2}\bigg),\,\,\,\, \text{for}\,\, n < m
\eea

\section{Debye mass approximated $\text{Im}~V$}
\label{appD}
In this case, once usually does not consider the general structure of the gluon propagator in presence of temperature and external magnetic field and instead incorporates the effect of the magnetic field solely through the modification in the Debye mass. Hence the imaginary part of potential in this case can be written as~\cite{Thakur:2013nia,Thakur:2020ifi}
		\bea
		\text{Im} V(r)&=&-\al T \phi_2(m_D \,r)-\frac{\sigma T}{m_D^2}\chi(m_D\, r),
		\label{imvgen}
		\eea
		where $m_D$ is the Debye screening mass. In order to calculate the Debye screening mass we have taken the static limit of the temporal component of the gluon self-energy i.e. $m_D^2=\Pi^{00}(\om\rightarrow 0, \bold p=0)$, where $\Pi^{00}(P)=\Pi^{00}_g(P)+\Pi^{00}_q(P)$.
First term in eq.~\ref{imvgen} comes from the Coulombic contribution whereas the second term is related to the string part of the Cornell potential.	
	
		The functions $\phi_2(x)$ and $\chi(x)$ are defined as,
		\bea
		\phi_2(x)&=& 2\int_0^\infty dz \frac{z}{(z^2+1)^2}\bigg(1-\frac{\sin(zx)}{z x}\bigg),\\
		\chi(x)&=&2\int_0^\infty dz \frac{1}{z(z^2+1)^2}\bigg(1-\frac{\sin(zx)}{z x}\bigg).
		\eea
	Both the functions $\phi_2(x)$ and $\chi(x)$ are monotonically increasing functions with $\phi_2(0)=0$ and $\chi(0)=1$. At large x, $\chi(x)$ is logarithmic ally divergent, whereas $\phi_2(\infty)=1$.

\bibliographystyle{apsrev4-1}
\bibliography{HQ.bib}

\end{document}